\documentclass[letterpaper, inpress]{jds} 

\jdsmonth{July}                 
\jdsyear{2020}                  
\jdsvolume{xx}                  
\jdsissue{xx}                   
\jdsdoi{xx.xxxx/xxxxxxxxx}      
\jdsreceived{April, 2020}       
\jdsaccepted{May, 2020}         

\usepackage{amsfonts,amsmath,amssymb,amsthm}
\usepackage{booktabs}
\usepackage{longtable}
\usepackage{float}
\usepackage{lipsum}
\usepackage[ruled,vlined]{algorithm2e}
\usepackage{caption}
\usepackage{subcaption}
\usepackage[acronym]{glossaries}

\theoremstyle{definition}
\newtheorem{definition}{Definition}[section]

\theoremstyle{remark}
\newtheorem*{remark}{Remark}

\title[A Review on GNN in finance]{A Review on Graph Neural Network Methods in Financial Applications}
\makenoidxglossaries
\newacronym{LSTM}{LSTM}{Long Short Term Memory}
\newacronym{GCN}{GCN}{Graph Convolutional Network}
\newacronym{RSR}{RSR}{Relational Stock Ranking}
\newacronym{MSE}{MSE}{Mean Squared Error}
\newacronym{IRR}{IRR}{Investment Return Ratio}
\newacronym{MRR}{MRR}{Mean Reciprocal Rank}
\newacronym{SFM}{SFM}{State Frequency Memory}
\newacronym{GBR}{GBR}{Graph-Based Ranking}
\newacronym{TRAN}{TRAN}{Time-aware graph Relational Attention Network}
\newacronym{LSTM-RGCN}{LSTM-RGCN}{LSTM Relational Graph Convolutional Network}
\newacronym{HAN}{HAN}{Hierarchical Attention Networks}
\newacronym{S-LSTM}{S-LSTM}{Sentence-state LSTM}
\newacronym{MAN-SF}{MAN-SF}{Multipronged Attention Network for Stock Forecasting}
\newacronym{MCC}{MCC}{Matthew’s Correlation Coefficient}
\newacronym{ARIMA}{ARIMA}{AutoRegressive Integrated Moving Average}
\newacronym{TSLDA}{TSLDA}{Topic Sentiment Latent Dirichlet Allocation}
\newacronym{VolTAGE}{VolTAGE}{Volatility forecasting via Text-Audio fusion with Graph convolution networks for Earnings calls}
\newacronym{MDRM}{MDRM}{Multimodal Deep Regression Model}
\newacronym{HTML}{HTML}{Hierarchical Transformer-based Multi-task Learning}
\newacronym{HAN1}{HAN}{Hybrid Attention Network}
\newacronym{CSI}{CSI}{China Securities Index}
\newacronym{SP500}{SP500}{Standard and Poor's 500}
\newacronym{NYSE}{NYSE}{New York Stock Exchange}
\newacronym{Nikkei  225}{Nikkei  225}{Nikkei Stock Average }
\newacronym{TPX}{TPX}{Tokyo Stock Price Index}
\newacronym{NASDAQ}{NASDAQ}{NASDAQ stock exchange}
\newacronym{AUC}{AUC}{Area Under the Curve}
\newacronym{SemiGNN}{SemiGNN}{Semisupervised
attentive Graph Neural Network}
\newacronym{KS}{KS}{Kolmogorov-Smirnov distance }
\newacronym{Xgboost}{Xgboost}{eXtreme gradient boosting}
\newacronym{LINE}{LINE}{Large-scale Information Network Embedding}
\newacronym{DGANN}{DGANN}{Dynamic
Graph-based Attention Neural Network}
\newacronym{GF}{GF}{Graph Factorization}
\newacronym{SEAL}{SEAL}{learning from Subgraphs, Embeddings and Attributes for Link prediction}
\newacronym{HGAR}{HGAR}{High-order Hraph Attention
Representation}
\newacronym{GRNN}{GRNN}{Graph Recurrent Neural Network}
\newacronym{precision@k}{Precision@k}{Precision of the top k nodes}
\newacronym{AANE}{AANE}{Accelerated Attributed Network Embedding}
\newacronym{SNE}{SNE}{attributed Social Network Embedding}
\newacronym{TRACER}{TRACER}{TempoRal Attention Contagion
chain Enhanced Rating model}
\newacronym{GBDT}{GBDT}{Gradient Boosting Decision Tree}
\newacronym{DNN}{DNN}{Deep Neural Network-based model}
\newacronym{ST-GNN}{ST-GNN}{Spatial-Temporal
aware Graph Neural Network}
\newacronym{STAR}{STAR}{Spatio-Temporal Attentive
Recurrent neural network}
\newacronym{AMG-DP}{AMG-DP}{Attributed Multiplex Graph based loan Default Prediction
approach}
\newacronym{MLP}{MLP}{Multi-Layer Perceptron}
\newacronym{MvMoE}{MvMoE}{Multi-view-aware
Mixture-of-Experts network}
\newacronym{GRC}{GRC}{Graph neural network with a Role-constrained Conditional random field}
\newacronym{SVM}{SVM}{Support Vector Machine}
\newacronym{GGNN}{GGNN}{Gated Graph Neural Network}
\newacronym{PMI}{PMI}{model based on Pairwise Mutual Information}
\newacronym{RNN}{RNN}{Recurrent Neural Network}
\newacronym{GAT}{GAT}{Graph Attention Network}
\newacronym{AutoGCN}{AutoGCN}{Auto-encoder based Graph Convolutional Networks}
\newacronym{DE}{DE}{Detection Expansion}
\newacronym{HGT}{HGT}{Heterogeneous Graph Transformer}
\newacronym{ASGCN}{ASGCN}{Adaptive Sampling Graph Convolutional Network}
\newacronym{mGCN}{mGCN}{modified version of Graph Convolutional Network}
\newacronym{HMGNN}{HMGNN}{Heterogeneous Mini-Graphs Neural Network}
\newacronym{CNN}{CNN}{Convolutional Neural Network}
\newacronym{GEM}{GEM}{Graph Embeddings for Malicious accounts}
\newacronym{DHGReg}{DHGReg}{Dynamic Heterogeneous Graph Neural Network}
\newacronym{GAL}{GAL}{Graph Anomaly Loss}
\newacronym{DOMINANT}{DOMINANT}{Deep anOMaly
detectIoN on Attributed NeTworks}
\newacronym{GraphSAGE}{GraphSAGE}{SAmple and aggreGatE}
\newacronym{HiGNN}{HiGNN}{Hierarchical bipartite Graph Neural Network}
\newacronym{Bi-HGNN}{Bi-HGNN}{Bipartite   Hierarchical bipartite Graph Neural Network}
\newacronym{DIN}{DIN}{Deep Interest Network}
\newacronym{GE}{GE}{Graph Embedding-based method}
\newacronym{Diffpool}{Diffpool}{Differentiable graph pooling}
\newacronym{GAS}{GAS}{GCN-based
Anti-Spam model}
\newacronym{GraphRfi}{GraphRfi}{GCN-based user Representation learning framework}
\newacronym{MAE}{MAE}{Mean Absolute Error}
\newacronym{RCF}{RCF}{Robust Collaborative Filtering model}
\newacronym{GCMC}{GCMC}{Graph Convolutional
Matrix Completion}
\newacronym{PMF}{PMF}{Probabilistic Matrix Factorization}
\newacronym{ICF}{ICF}{Item-based Collaborative Filtering}
\newacronym{MF}{MF}{Matrix Factorization}
\newacronym{RMSE}{RMSE}{Root Mean Squared Error}
\newacronym{GCNEXT}{GCNEXT}{Graph Convolutional Network with Expended Balance Theory}
\newacronym{RGCN}{RGCN}{Relational Graph Convolutional Network}
\newacronym{SGCN}{SGCN}{Signed Graph Convolutional Network}
\newacronym{SIDE}{SIDE}{SIgned Directed network Embedding}
\newacronym{CARE-GNN}{CARE-GNN}{CAmouflage-REsistant GNN}
\newacronym{RF}{RF}{Random Forest}
\newacronym{LR}{LR}{Linear Regression}
\newacronym{DW}{DW}{DeepWalk}

\author[1]{Jianian Wang }
\author[1]{Sheng Zhang}
\author[2]{Yanghua Xiao \thanks{Corresponding author. Email: shawyh@fudan.edu.cn}}
\author[1]{Rui Song \footnote{Corresponding author. Email: rsong@ncsu.edu}}

\affil[1]{Department of Statistics, North Carolina State University, Raleigh, United States}
\affil[2]{School of Computer Science, Fudan University, Shanghai, China}

\begin{document}

\maketitle

\begin{abstract}
  {\color{black}With multiple components and relations, financial data are often presented as graph data, since it could represent both the individual features and the complicated relations.  Due to the complexity and volatility of the financial market, the graph constructed on the financial data is often heterogeneous or time-varying, which imposes challenges on modeling technology. Among the graph modeling technologies, graph neural network (GNN) models are able to handle the complex graph structure and achieve great performance and thus could be used to solve financial tasks}. In this work, we provide a comprehensive review of GNN models in recent financial context. We first categorize the commonly-used financial graphs and summarize the feature processing step for each node. Then we summarize the GNN methodology for each graph type, application in each area, and propose some potential research areas. \@
\end{abstract}

\section{Introduction}%
\label{sec:intro}
As the data collection techniques grow, graph data are commonly collected in many areas including social sciences, transportation systems, chemistry, and physics \citep{Wu_review}. Representing complex relational data, graph data contain both the individual node information and the structural information. Recently, there are growing interests in developing machine-learning methods to model the graph data of various domains. {\color{black}Among them, graph neural network (GNN) methods could achieve great performance on various tasks, including node classification, edge prediction, and graph classification} \citep{GCN,Zhang_link, GIN}. Performing node aggregation and updates, graph neural network models extend the deep learning methodology to graphs and are gaining popularity. 

\begin{figure}
    \centering
    \includegraphics[width=\textwidth]{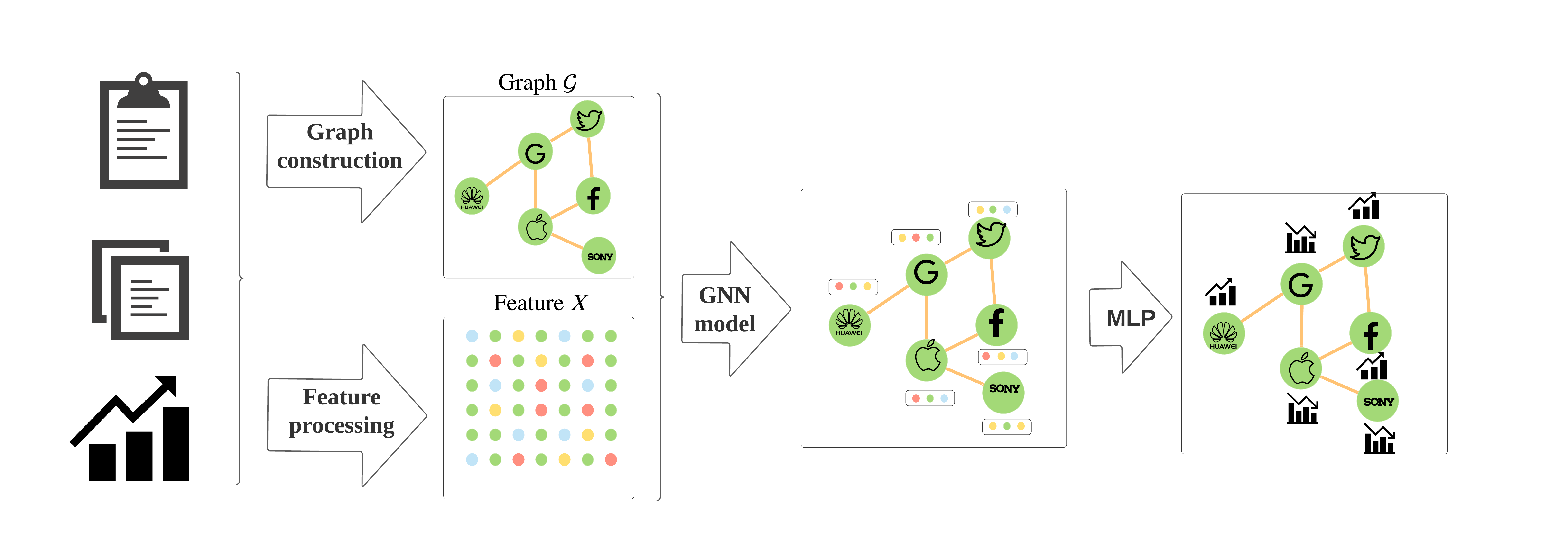}
    \caption{Workflow for stock movement prediction task using GNN methodology. The graph construction and feature processing steps present stock information in a graph and a feature matrix, which is then used as the input for the GNN model. {\color{black} In the graph, nodes are connected if there exist some relationships between stocks, such as supplier, competitor, shake-holder, etc.} A multi-layer perception layer (MLP) is used to output the price prediction result. }
    \label{workflow}
\end{figure}

{\color{black}A financial system is a complex system with many components and sophisticated relations, which may be frequently updated.  To represent the relational data in the financial domain, graphs are commonly constructed, including the transaction network \citep{Weber}, user-item review graph \citep{Dou}, and stock relation graph \citep{Feng}. By converting the financial task into a node classification task, GNN methods are commonly utilized since it performs well among graph modeling methods \citep{Liu_19}.} For instance, GNN could be utilized in a stock prediction task, by formulating it as a node classification task, where each node represents a stock and edges represent relations between companies. Figure \ref{workflow} demonstrates the workflow of a stock prediction task using GNN methods. However, the complex nature of financial systems may result in multiple data sources and complicated graph structures, which imposes challenges on feature processing, graph construction, and graph neural network modeling. Represented as numerical sequences or textual information, financial data need to be processed with caution to keep the temporal pattern or semantic meanings. Also, the multi-facet nature of financial relations make it hard to construct a graph to capture the relations.
{\color{black}Moreover, the financial-related graph is often heterogeneous or time-varying, which impose challenges on existing graph neural network models. What's more, to reflect some financial patterns (e.g. device aggregation pattern, see Section \ref{fraud} for details), GNN methods may need to be modified such as changing losses and adding additional layers.} Since financial systems process unique characteristics and receive great attention, it is of significant importance to discuss and summary the GNN methodology developed for financial tasks.

There are several recent reviews on graph neural networks. Among them, \citet{Wu_review} present a comprehensive review on graph neural networks and categorize the GNNs into four categories: recurrent graph neural networks, convolutional graph neural networks, graph auto-encoders, and spatial-temporal graph neural networks. \citet{Zhou_review} provide a taxonomy on GNN models based on graph type, training methods, and propagation steps. There is also literature focusing on limited types of GNNs.  \citet{Zhang_GCN_review} focus on graph convolutional networks (GCN) and introduce two taxonomies to group the existing GCNs. \citet{Lee_review} survey the literature on graph attention models and provided detailed examples on each type of method. However, the aforementioned reviews focus on the general methodology and provide little details for applications, seldom mentioning the financial application. Without covering GNN models developed based on financial contexts, the reviewed models may not be applicable to financial tasks due to the complexity of financial data. On the other side, review papers focusing on the financial domain haven't covered GNN methodologies in detail yet. \citet{Ozbayoglu} summarize the machine learning and deep learning models in the financial field, without mentioning the GNN methodologies. \citet{Huang_review} survey the financial deep learning models in the finance and bank industry, and the GNN models are not covered. \citet{Jiang_review} review stock prediction-related machine-learning mythologies and mention GNN models very briefly. In summary, existing GNN surveys focus on modeling methodology and do not emphasize the financial application of GNN methods, while surveys on financial applications don't cover the GNN models in detail. To fill the gap, in this survey, we provide a systematic and comprehensive review of graph neural network methods in the financial application.

In this paper, we present a thorough survey on graph neural network models with financial application. 
We provide a comprehensive review of graph neural networks and summarize the corresponding methods. This survey has contributions as follows.

\begin{itemize}[leftmargin=1cm]
    \item We systemically categorize the commonly-used financial graphs based on graph characteristics and provide a thorough list of graphs. Graphs are categorized into five groups: homogeneous graph, directed graph, bipartite graph, multi-relation graph, and dynamic graph. We also present the GNN models according to their graph types, so that this review could serve as a guide for implementing GNNs on real-life datasets.
    \item We provide a comprehensive list of financial applications that GNN methods are applied. We categorized the applications into five categories: stock movement prediction, loan default risk prediction, recommender system of e-commerce, fraud detection, and event prediction.
    \item We summarize various aspects of information for each application, including features, graphs, GNN models, and available codes. A GitHub\footnote{Github link: https://github.com/jackieD14/Graph-models-in-finance-application} page is built to document the collection of information. This work could be considered as a resource to understand, implement and develop GNN models on multiple financial tasks. 
    \item We identify five challenges and discuss the recent progress. We also suggest future directions for these problems.
\end{itemize}

The rest of the paper is organized as follows. Section \ref{session2} classifies financial graphs into different categories based on its characteristics. Section \ref{session3} summarizes the commonly-used feature processing techniques for each node in the graph. Section \ref{session4} presents the GNN methodology used for each graph type. Section \ref{session5} provides a collection of application areas.  Section \ref{session6} proposes some challenges that could be future directions of research.

\section{Graph categorization}
\label{session2}
When preparing the data, how to construct the graph to represent the structural information is essential and the type for the constructed graph could determine the follow-up modeling methodology. In this section, we present the categorization of the graph based on its construction methods and graph types. Table \ref{table_summary} presents a comprehensive list of graphs for financial tasks. 
\subsection{Graph-related definition}
In this section, we provide some graph-related definitions for better understanding of this article.
\theoremstyle{definition}
\begin{definition}[Graph]
	A graph $\mathcal{G}$ is defined by a pair: $\mathcal{G}=(\mathcal{V},\mathcal{E})$, where $\mathcal{V}=\{v_1,...,v_n\}$ is a set of $n$ nodes and $\mathcal{E}$ is a set of edges, where $e_{ij}=(v_i,v_j)\in \mathcal{E}$ denotes an edge joining node $v_i$ and node $v_j$.
\end{definition}
\begin{definition}[Adjacency matrix]
	An adjacency matrix $\textbf{A}$ is an $n \times n$ matrix, where $\textbf{A}_{ij}$ represents the connection status between node $v_i$ and node $v_j$. For an unweighted graph, the adjacency matrix could be an binary matrix where $\textbf{A}_{ij}=1$ if $e_{ij}\in \mathcal{E}$ and 0 otherwise.
\end{definition}
\begin{definition}[Undirected graph and Directed graph]
	A undirected graph is a graph where the edges are undirected.
	A directed graph is a graph where the edges have orientations. $e_{ij}=(v_i,v_j)\in \mathcal{E}$ denotes an edge pointing from node $v_i$ to node $v_j$.
\end{definition}
\begin{remark}
	Undirected graph has a symmetric adjacency matrix, i.e., $\textbf{A}_{ij}=\textbf{A}_{ji}$.
\end{remark}

\begin{definition}[Bipartite graph]
	A Bipartite graph is a graph whose nodes could be divided into two non-empty and disjoint sets $\mathcal{U},\mathcal{W}$, such that every edge connects a node in $\mathcal{U}$ and a node in $\mathcal{W}$.
\end{definition}
\begin{definition}[Homogeneous graph and Heterogeneous graph] In a graph $\mathcal{G} = (\mathcal{V},\mathcal{E})$, we can assign a type to each node and edge; in this case, the graph is denoted as $\mathcal{G}=(\mathcal{V},\mathcal{E}, \mathcal{A}, \mathcal{R})$, where each node $v_i\in \mathcal{V}$ is associated with its type $a_i\in \mathcal{A}$, and each edge $ e_{ij}\in \mathcal{E}$ is associated with its type $r_{ij} \in \mathcal{R}$. A homogeneous graph is a graph whose nodes are of the same type and edges are of the same type. Otherwise, the graph is heterogeneous. 
\end{definition}
\begin{definition}[Multi-relation graph] {\color{black}A Multi-relation graph is a graph where edges have different types.}
\end{definition}
\begin{definition}[Dynamic graph] A dynamic graph is defined as a sequence of  graphs $\mathcal{G}^{seq}=\{\mathcal{G}_1,...,\mathcal{G}_T\}$, where $ \mathcal{G}_i=(\mathcal{V}_i,\mathcal{E}_i), \text{for } i=1,...,T$, where $\mathcal{V}_i,\mathcal{E}_i$ are the set of nodes and edges for $i^{th}$ graph in the sequence respectively.
\end{definition}

\subsection{Graph categorization by construction methods}
In this section, we elaborate on frequently-used graph construction methods so that researchers could better understand, choose and construct graph data.

\subsubsection{Data-based construction}
Some types of data could be naturally represented as a graph since they contain relations among data objects. For instance, \citet{Wang} construct a user-relationship graph, where users are linked by an edge if they are labeled classmates, friends, or workmates in the data. \citet{Liu_18} construct an account-device network, where nodes are either accounts or devices.
Edges connect an account node to a device node if the account has activities on the device. This type of construction method is based on the nature of data and could be used on data where relations are clearly defined.

\subsubsection{Knowledge-based construction}
Sometimes, data may not contain relational information, but relations could be found in knowledge bases. A knowledge base is a collection of descriptive data and contains numerous entities and their relations. For example, Wikidata \citep{Wikidata} is one of the largest open-domain knowledge bases which provides support for Wikipedia, an online encyclopedia. A graph could be built utilizing the relations extracted from knowledge bases. In order to predict stock movement, \citet{Feng} extract company relations from Wikidata, such as supplier, provider, partner, etc, and constructed a company-based relation network. This type of graph-construction method brings new information into the graph by utilizing the knowledge bases, which may improve modeling performance. However, it may take some effort to process the complicated data structure and the massive amount of information of knowledge bases.

\subsubsection{Similarity-based construction}
There also exist cases that neither the data nor knowledge bases contain relations, but there may be some hidden relationships in the data. To mine the underlying relationship, a commonly-used approach is to calculate a similarity measure of the features for different observations and construct relations if the similarity value is greater than a threshold. For instance, \citet{Li} construct a stock correlation graph based on the cosine similarity of stocks' historic market price. Two stocks are then connected if the absolute value of their correlation is larger than a threshold. This type of construction method could be easily implemented and understood. However, in this type of construction, feature information is represented both in the graph adjacency matrix and the feature matrix. The overlapping information may lead to doubts that whether the graph representation is still necessary. Also, how to set the threshold is an issue, and justification may be needed for the selected similarity threshold value.

\subsection{Graph categorization by graph types}
\begin{figure}[H]
    \centering
    \includegraphics[width=0.8\textwidth]{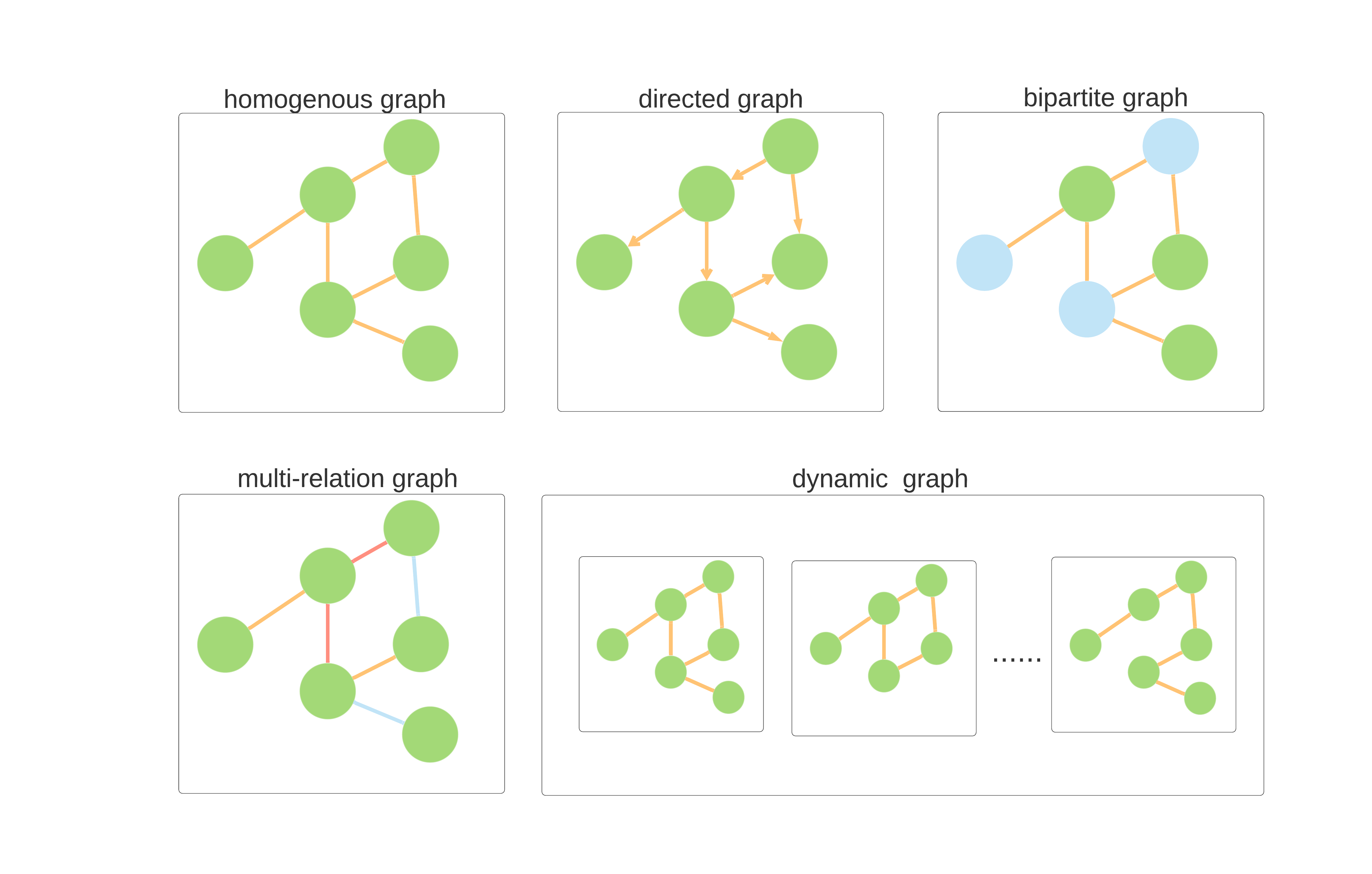}
    \caption{Graph categorization based on graph characteristics. Each color of the circle represents a node type and each color of the line represents an edge type. Arrows represent directed edges. A homogeneous graph is a graph with one type of node and one type of edge. A directed graph is a graph with directed edges. A bipartite graph is a graph with two types of nodes and edges only exist between nodes of different types. A multi-relation graph has edges with different types. A dynamic graph is a sequence of graphs. }
    \label{graph_type}
\end{figure}

 {\color{black}In this section, we categorize graphs into five categories based on their characteristics and provide examples in the financial context. Since different types of graphs may impose various challenges on modeling technology, we discuss GNN methods for each graph type in section \ref{session4} to provide solutions respectively}. Figure \ref{graph_type} provides visualization for each graph type:  homogeneous graph, directed graph, bipartite graph, multi-relation graph, and dynamic graph. It is also worth mentioning that a graph may be categorized into multiple types.
 
\subsubsection{Homogeneous graph}
A homogeneous graph is a graph with one type of node and one type of edge. For instance,  \citet{Liou} construct a financial news co-occurrence graph, where two companies are connected if they are tagged in the same news articles. \citet{Li_laundry} build a transaction network, where nodes are accounts and are connected when there exist transactions between them. This type of graph has a relatively simple structure and the majority of the GNN methods could be applied to model this type of graph.

\subsubsection{Directed graph}
A directed graph is a graph where edges have orientations. For instance, \citet{Cheng_network_loan} construct a guarantee network where nodes are the companies and edges represent the guarantee relationship. Since the guarantor has the obligation to pay the debt for the borrower, but not the other way round, this type of guarantee relationship is one-sided and could be represented in a directed edge. In general, a directed graph may have an asymmetric adjacency matrix and thus cannot be semi-definite. Since some GNN mythologies are developed for semi-definite adjacency matrices, they may not be suitable for the directed graph.

\subsubsection{Bipartite graph}
 A bipartite graph is a graph with two types of nodes and edges only exist between nodes of different types. For instance, \citet{Liu_18} construct an account-device network in a fraud detection task. Nodes could be either accounts or devices, with edges connecting them if the account has activities on the device. \citet{Li_spam} extract a user-item network using the rating data, where nodes could be either a user or an item. An edge exists if a user has rated the item. A bipartite graph is commonly used, when data can be divided into two groups and the interaction between two groups matters. A bipartite graph could be seen as a specific case of a multi-relation graph discussed in the next session and GNN methods developed for the multi-relation graph are applicable on a bipartite graph as well.

\subsubsection{Multi-relation graph}
Sometimes, edges may have multiple types to represent different relations between nodes. For instance, \citet{Wang} construct a user-relation graph, where nodes are the users of an e-commerce platform. There are multiple edge types representing various relationships including friendship, workmates and classmates. \cite{Dou} build a review graph where users are represented as nodes in the graph. Three types of relations between users are defined to capture their behavioral patterns: reviewing the same product, having the same star rating, and having similar texts. With multiple edge types, this type of graph contains more information about the relationship between nodes, and thus how to capture this information is critical when developing the GNN methodology.

\subsubsection{Dynamic graph} A dynamic graph is a sequence of graphs , where each graph could have an adjacency matrix and feature matrix. For example, \citet{Cheng_IJCAI} construct a temporal guarantee network, representing the guarantee relationship in each time step.  This type of graph is commonly used to represent the changes in both relations and features, as time goes. Since both nodes and edges could appear and disappear, it is hard to perform some graph operations that require fixed dimensions of matrices. Thus, capturing the dynamically of the graph is challenging and requires a more sophisticated methodology. 
\begin{longtable}{p{.4\textwidth}p{.15\textwidth}p{.1\textwidth}p{.1\textwidth}p{.2\textwidth}}
		\toprule
		Graph & Construction method \footnote{\textit{Knowledge, Similarity, Data} denotes knowledge-based, similarity-based and data-based construction respectively.}& Graph type \footnote{\textit{Multi} denotes multi-relation graph and \textit{Homo} denotes homogeneous graph.} & Application \footnote{\textit{Stock} denotes stock movement prediction. \textit{Loan} denotes loan default risk prediction. \textit{E-comm} denotes recommender system of e-commerce. \textit{Fraud} denotes fraud detection.}\footnote{ {\color{black}Details for financial applications could be found in Section \ref{session5}}.}  & Reference\\
		\midrule
		Sector-industry stock relation network & Knowledge  & Multi & Stock& \citet{Feng}\\
		Wiki company-based relation network & Knowledge  & Multi & Stock& \citet{Feng,Sawhney,Ying}\\
		Supplier, customer, partner and shareholder  relation graph &  Knowledge & Multi & Stock& \citet{Matsunaga}\\
		Corporation shareholder network & Knowledge  & Homo & Stock& \citet{Chen}\\
		Stock correlation graph & Similarity  & Multi & Stock& \citet{Li}\\ 
		Stock earning call graph & Knowledge  & Bipartite & Stock& \citet{Sawhney_audio}\\
		News co-occurrence graph & Similarity  & Homo & Stock& \citet{Liou}\\
		User-relation graph &Data & Homo & Loan & \citet{Wang}\\
		User-app graph &Data & Bipartite & Loan & \citet{Wang}\\
		User-nickname graph &Data & Bipartite & Loan & \citet{Wang}\\
		User-address graph &Data & Bipartite & Loan & \citet{Wang}\\
		Guarantee network &Data & Directed & Loan & \citet{Cheng_network_loan,Cheng_KDD}\\
		Temporal guarantee network &Data & Dynamic & Loan & \citet{Cheng_IJCAI}\\
		Temporal small business entrepreneur network &Data & Dynamic & Loan & \citet{Yang}\\
		Alipay user and applet graph &Data & Dynamic & Loan & \citet{Hu_Loan}\\
		User relationship graph &Data & Multi & Loan & \citet{Liang_WSDM}\\
		Auto relation network &Data & Multi & Loan & \citet{Xu}\\
		Loan application event graph &Data & Directed & Loan & \citet{Harl}\\
		Borrower relations' network & Similarity & Directed & Loan & \citet{Lee}\\
		Xianyu comment graph& Similarity  & Homo & E-comm & \citet{Li_spam}\\
		Yelp review network &Data & Bipartite & E-comm & \citet{Zhang_Fraud,Dou}\\
		Amazon review network &Data & Bipartite & E-comm & \citet{Zhang_Fraud,Dou,Kudo}\\
		E-commerce user-item network &Data & Bipartite & E-comm & \citet{Li_hierarchical}\\
		Taobao user-item network &Data & Bipartite & E-comm & \citet{Li_bipartite}\\
		Device sharing graph &Data & Bipartite & Fraud & \citet{Liang}\\
		JD Finance anti-fraud graph &Data & Multi & Fraud  & \citet{Lv}\\
		Transaction records graph &Data & Multi & Fraud  & \citet{Rao1}\\
		Iqiyi user network  &Data & Multi & Fraud  & \citet{Zhu}\\
		CMU simulated user activity network & Similarity  & Homo & Fraud & \citet{Jiang_anomaly}\\
		Alipay one-month account-device network &Data & Bipartite & Fraud & \citet{Liu_18}\\
		Alipay one-week account-device network &Data & Bipartite & Fraud & \citet{Liu_19}\\
		Account-registration graph &Data & Dynamic & Fraud & \citet{Rao}\\
		Bitcoin-alpha graph &Data & Directed & Fraud & \citet{Zhao_loss}\\
		\bottomrule\\
\caption{Summary of financial-related graphs}
\label{table_summary}
\end{longtable}

\section{Feature processing}
\label{session3}
{\color{black}With diverse data sources in the financial field, node features are commonly formatted as sequential numerical features or textual information. These data formats impose challenges on the feature processing step since GNN methods could not be directly applied to these data formats. In this section, we summarize the commonly-used feature processing techniques and how they solve these challenges.} 

\begin{figure}[H]
     \centering
     \begin{subfigure}[b]{0.55\textwidth}
         \centering
         \includegraphics[width=\textwidth]{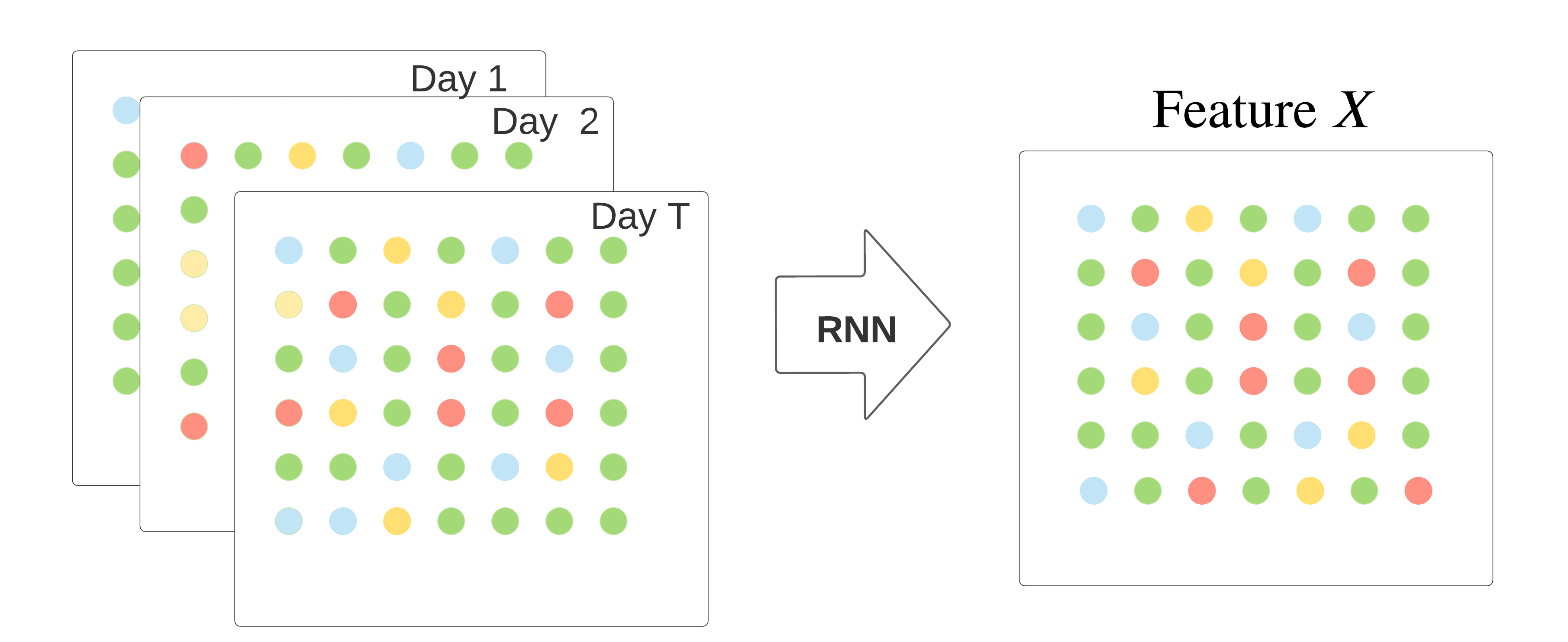}
         \caption{Sequential numerical features}
     \end{subfigure}
     \hfill
     \begin{subfigure}[b]{0.4\textwidth}
         \centering
         \includegraphics[width=\textwidth]{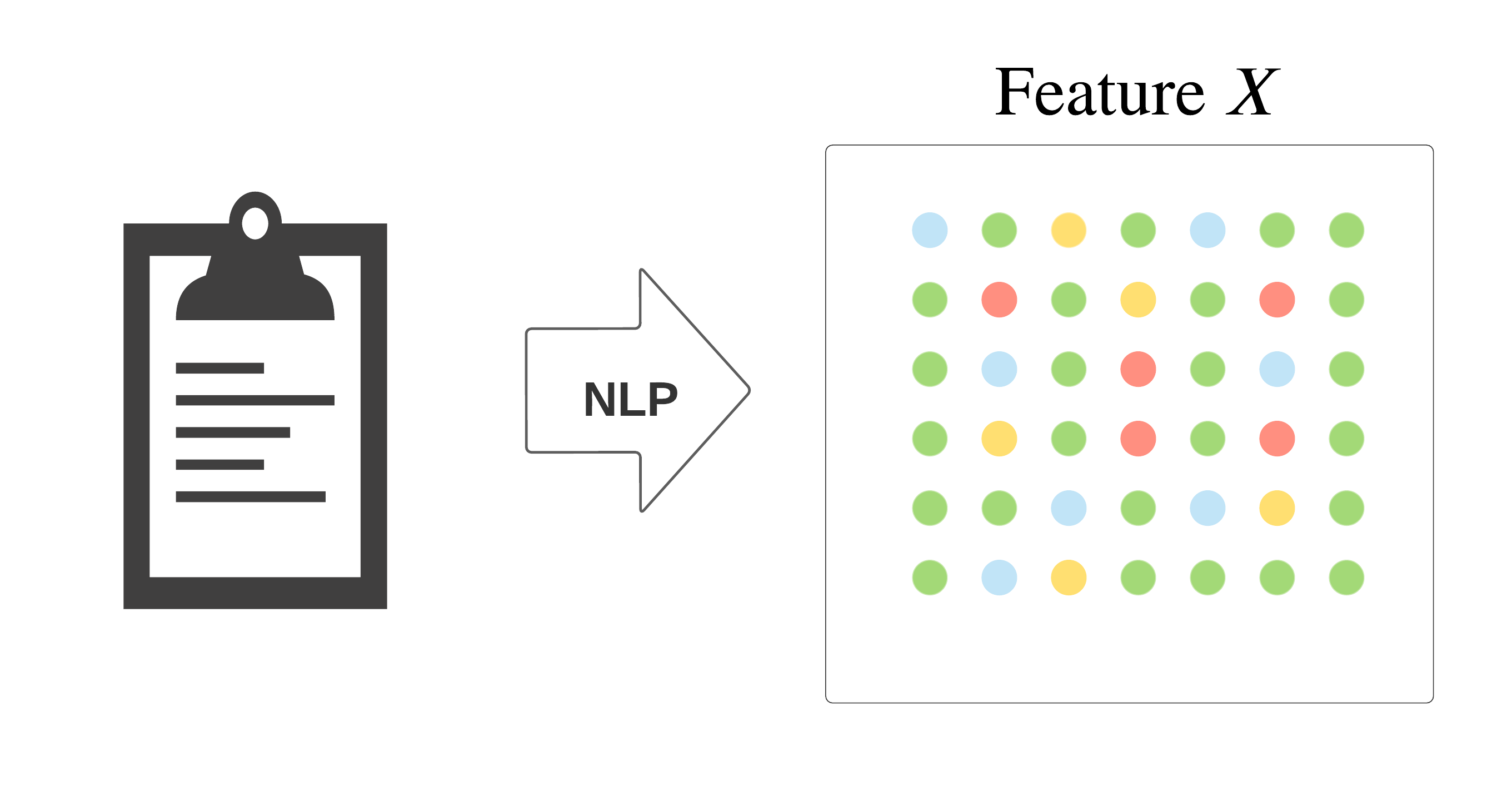}
         \caption{Textual information}
     \end{subfigure}
        \caption{Feature processing for sequential features and textual information. For sequential numerical features, recurrent neural network (RNN) based approaches are commonly used to capture the temporal dependencies. For text features, it is often processed utilizing natural language processing (NLP) methods including word embedding, sentence embedding, and language models, to convert the unstructured data to structured ones. }
        \label{feature}
\end{figure}
\subsection{Sequential numerical data}
Updating information as time goes, the financial industry has a rich source of time-series data. Indexed using timestamps, features could be seen as sequential which requires appropriate modeling. Consider the feature matrix at time $s$, $X^{s}\in \mathbb{R}^{n\times p \times l}$, where $n$ is the number of nodes, $p$ is the dimension of features at each time point, $l$ is the length of the sequence. For example, in a stock prediction task, we have a stock relation graph, with $n$ stocks as nodes. The feature matrix $X^s$ represents the $p$ features in the past $l$ days from time $s$ for these $n$ stocks. To encode the numerical sequence for each node, a recurrent neural network (RNN) is frequently used due to its superior performance in predicting time-series data. The literature could be summarized into two lines of work, long short-term memory (LSTM) based approach and gated recurrent unit (GRU) based approach.

\subsubsection{LSTM based approach}
As a special form of recurrent neural network, long short-term memory(LSTM) \citep{LSTM} is capable to capture the long-term dependencies and avoid the vanishing gradient problem. Using memory cells and gate units, it has the following expression:
\begin{align*}
	f_t&=\sigma (W_f x_t+U_fh_{t-1}+b_f),\\
	i_t&=\sigma (W_i x_t+U_ih_{t-1}+b_i),\\
	o_t&=\sigma (W_o x_t+U_oh_{t-1}+b_o),\\
	\tilde{c}_t&=\text{tanh} (W_c x_t+U_ch_{t-1}+b_c),\\
	c_t&=f_c \circ c_{t-1}+i_t\circ \tilde{c}_{t-1},\\
	h_t&=o_t \circ \text{tanh}(c_t),
\end{align*}
where $x_t\in \mathbb{R}^{D}$ is the input vector at time $t$ and $D$ is the number of features, $f_t,i_t,o_t,\tilde{c}_t,c_t,h_t$ denotes the forget gate, output gate, cell input, cell state and hidden state vectors respectively, $W_f,W_i$, $W_o,W_c$ $,U_f, U_i, U_o, U_c$ are trainable weight matrices and $b_f,b_i,b_o,b_c$ are trainable bias vectors, $\sigma(\cdot)$represents the sigmoid activation function, and $\circ$ denotes the element-wise product. The hidden state of the LSTM on day $t$ is denoted by: $h_t=LSTM(x_t, h_{t-1}), s-l\leq t\leq s$.

Since the LSTM updates the hidden state to capture the structural information, a common approach to encode the historical data is generating sequential embedding $E^s$ using the last hidden state of LSTM, $E^s=\text{LSTM}(X^s)\in \mathbb{R}^{n \times u}$, where $u$ is the dimension of the output feature. Then the encoded pricing information is used as input to the graph neural network. For instance,  \citet{Chen} used the generated sequential embedding $E^s$ as the input feature matrix for graph convolutional network and \citet{Feng} utilized  $E^s$ as the input feature matrix for their proposed temporal graph convolutional layer. Using the last hidden state as an input feature, this type of method could capture the information in the past days while having an appropriate format to feed into the GNN model.

\subsubsection{GRU based approach}
Gated recurrent unit (GRU) \citep{GRU} is another variant of RNN models. It also applies gating mechanism and has fewer parameters. It has the following structure:
\begin{align*}
r_t&=\sigma (W_r x_t+U_rh_{t-1}+b_r),\\
z_t&=\sigma (W_z x_t+U_zh_{t-1}+b_z),\\
\tilde{h}_t&=\text{tanh} (W_h x_t+U_h (r_t \circ h_{t-1})+b_h),\\
h_t&=(1-z_t) \circ h_{t-1}+z_t \circ \tilde{h}_t,
\end{align*}
where $x_t\in \mathbb{R}^{D}$ is the input vector at time $t$ for stock $i$ and $D$ is the number of features, $r_t,z_t,\tilde{h}_t, h_t$ denotes the reset gate, update gate, candidate activation and hidden state vectors respectively, $W_r,W_z,W_h$ $,U_r,U_z,U_h$ are trainable weight matrices and $b_r,b_z,b_h$ are trainable bias vectors, $\sigma(\cdot)$ represents the sigmoid activation function, and $\circ$ denotes the element-wise product. The hidden states of the GRU on day $t$ is denoted by: $h_t=\text{GRU}(x_t, h_{t-1})$.

Utilizing GRU to encode the past numerical information, we could obtain the hidden state for each day. Since past days' impact on the current-day representation may differ, an attention mechanism is frequently used to assign weights differently. For instance, \citet{Sawhney} use an additive attention mechanism to aggregate the hidden states across time. \citet{Cheng_KDD} utilize the concatenated attention method to incorporate different importance of time. The attention mechanism aggregates the hidden states of past days and assigns different weights across time. To get the feature representation at time $s$, it rewards the influential days when aggregating the hidden states from time $s-l$ to time $s$, and thus take temporal dependencies into account. The obtained node representation is then used as a feature in the graph neural network model. 




\subsection{Textual information}
In the financial industry, a large amount of information is of textual form, including financial news, financial statements, and customer reviews. How to translate the texts into vector representation while preserving semantic information, is essential. The following section summarizes the commonly-used natural language processing (NLP) methodology to convert the unstructured texts into a vector form.
\subsubsection{Word embedding and sentence embedding}
Word embedding methods are widely used to represent a word as a fixed-length vector. Then, to learn the sentence representation, a recurrent neural network model is often utilized to capture local semantic information. For instance, to embed the news headlines, \citet{Li} encode the word as word embedding using GloVe \citep{Glove}. Then LSTM method is applied with an attention mechanism to create the sentence representation. However, since a word may have multiple meanings, word embedding methods may cause problems by assigning the same vector to words with different meanings. Thus, there are also approaches to embed at a sentence level in order to alleviate the problem. For instance, \citet{Sawhney} generate sentence-level embedding for each Tweet using Universal sentence encoders \citep{USE}.

\subsubsection{Language model}
Instead of focusing on generating vectors for words, language models focus on capturing the pattern of languages to predict the word based on its surrounding words. Taking the contexts into account, language models achieve the state of art performance on many NLP tasks and are widely used in the literature.  For instance, \citet{Liou} use bidirectional encoder representations from
transformers (BERT) model \citep{BERT} to encode the entire news article and generated a news embedding. Without modifying the model architecture, the pre-trained BERT model is able to be fine-tuned and produce state-of-art performance. For example, FinBERT \citep{Finbert}, a BERT model pre-trained specific to the financial domain, is utilized to encode the text scripts in company earning calls \citep{Sawhney_audio}.

\section{Graph neural network models}
Proposed by \citet{Gori}, a graph neural network is a neural network model capable of processing graphs. Unlike network embedding methods whose major aim is to generate a vector to represent each node, graph neural network models are designed for a variety of tasks, including node classification, edge prediction, and graph classification. Due to its wide application and superior performance, graph neural network models have drawn great attention. In this section, we present the commonly-used GNN models for each type of graphs, since the methodology may vary with different graph characteristics. In the supplementary materials, we present a figure demonstrating the major GNN methodology used for each graph type. 
\label{session4}

\subsection{Homogeneous graph}	
Proposed by \citet{GCN}, graph convolutional network (GCN) is a widely-used graph neural network model and could encode both local graph structure and node features. Extending the convolution concepts to graphs, graph convolution could be seen as message passing and information propagation. Aggregating neighbors' feature information, graph convolutional networks could be represented with the following layer-wise propagation rule:
\begin{align*}
H^{l+1}=\sigma(\tilde{D}^{-\frac{1}{2}}\tilde{A}\tilde{D}^{-\frac{1}{2}}H^{l}W^{l}),
\end{align*}
where $\tilde{A}$ is the adjacency matrix with added self-connections, $\tilde{D}=\text{diag}(\sum_j \tilde{A}_{ij})$ $W^{l}$ is trainable weight matrix of $l^{th}$ layer, $H^{l}$ is the node hidden feature matrix in the $l^{th}$ layer and $\sigma(\cdot)$ is the activation function.
With its relatively simple model structure and great performance, the GCN model is often used as a benchmark method to compare with. Among the reviewed literature, over half of them have applied the GCN method as a benchmark method.

While GCN equally treats the neighbors of the target node, it often occurs that some neighbors may be more influential than others. Considering various impacts of the neighbor nodes, \citet{GAT} propose graph attention networks (GAT) and it is able to assign different weights to nodes in the same neighborhood as follows:
\begin{align*}
h_i^{l+1}&=\sigma(\sum_{j\in \mathcal{N}_i}\alpha_{ij}^{l}W^{l}h_j^{l}),\\
\alpha_{ij}^l&=\frac{\exp(\text{LReLU}(a^T[W^{l}h_i^{l}\Vert W^{l}h_j^{l}] ]))}{\sum_{k\in \mathcal{N}_i}\exp(\text{LReLU}(a^T[W^{l}h_i^{l}\Vert W^{l}h_k^{l}]},
\end{align*}
where $h^{l}$ is the hidden feature vector for node $i$ in the $l^{th}$ layer, $W^{l}$ is trainable weight matrix, $a$ is a learnable vector, $\mathcal{N}_i$ is the neighborhoods of node $i$, $\alpha_{ij}^l$ represents attention coefficient of node $j$ to $i$ at $l^{th}$ layer, $\sigma(\cdot)$ is the activation function, $\Vert$ denotes vector concatenation, and $\text{LReLU}$ denotes the leaky ReLU activation function. It is also worth mentioning that, since GAT is able to learn the weights of the neighboring node, we could interpret the learned attention weights as a relative importance measure, to better understand the model. Similar to GCN, GAT is also often used as a benchmark method in the reviewed papers with about 40\% coverage.

\subsection{Directed graph}
A undirected graph has a symmetric adjacency matrix that guarantees a semi-definite Laplacian matrix, which lays the foundation for applying GCN. The directed graph, on the other hand, may has asymmetric adjacency matrix and could be handled by spatial-based GNN methods \citep{Wu_review}, such as GAT. In practice, there is not much work developing methodologies for directed graph, since it could be processed by spatial-based GNN methods or making the adjacency matrix symmetric. However, there is also a line of work developing GNN methodology to predict sequences of events represented as a directed graph.

A sequence of events could be naturally formed as a directed graph, where each node is one type of event and an edge points from one event to the following one. Given the graph, which could be seen as a partial sequence of events, how to encode the features and predict the rest of the sequence is a challenge. The aforementioned GCN and GAT models aim at representation learning and are used to produce a single output instead of outputting a sequence. To fill the gap, \citet{GGNN} propose a gated graph neural network (GGNN) that could produce sequential outputs. It applies gated recurrent unit (GRU) as a recurrent function and is constructed as follows:
\begin{align*}
a_t&=A^Th_{t-1}+b,\\
r_t&=\sigma (W_r a_t+U_rh_{t-1}),\\
z_t&=\sigma (W_z a_t+U_zh_{t-1}),\\
\tilde{h}_t&=\text{tanh} (W_h a_t+U_h (r_t \circ h_{t-1})),\\
h_t&=(1-z_t) \circ h_{t-1}+z_t \circ \tilde{h}_t,
\end{align*}
where $h_t$ is the updated event representation at $t^{th}$ step, $a_t$ contains information transferred from both directions' edges, $r_t,z_t,\tilde{h}_t$ denotes the reset gate, update gate, and candidate activation vectors at $t^{th}$ respectively, $W_r,W_z,W_h$ $,U_r,U_z,U_h$ are trainable weight matrices, $b$ is a trainable bias vector,  $\sigma(\cdot)$ is the sigmoid function, $\circ$ denotes the element-wise product, and $tanh$ denotes the hyperbolic tangent function. Incorporating the adjacency matrix $A$, GGNN aggregates the structural information in every propagation step. Unrolling the recurrence function to a fixed number, the GGNN ensures convergence without constraining the parameters.

\subsection{Bipartite graph}
\label{Bipartite}
	The aforementioned methods focus on homogeneous graphs whose nodes and edges are all of one type. In real-life applications, graphs could be heterogeneous. As a running example, in a spam detection task, we could have a
	user-item network, where nodes are either users or items.
	An edge $e_{ij}$ denotes that user $i$ has rated on item $j$. This type of graph is well known as a bipartite graph $\mathcal{G}$ with the following notation: $\mathcal{G}=(\mathcal{V},\mathcal{E})$, where  $\mathcal{V}=\{U,W\}$ is a set of nodes and could be divided into two non-empty and disjoint sets $\mathcal{U,W}$. $\mathcal{E}$ is a set of edges, where every edge joins a node in $\mathcal{U}$ and a node in $\mathcal{W}$.
	
    Based on the characteristics of a bipartite graph, the commonly-used methodologies could be categorized into a framework as follows. In each iteration, the edge information $h_{uw}^{l}$ is updated aggregating its previous hidden state $h_{uw}^{l-1}$ and hidden states of the two nodes it links ($h_{u}^{l-1}$, $h_{w}^{l-1}$), as shown in equation \eqref{edge-update}. At each iteration, a user node $u$ aggregates information from its rated items $h_{\mathcal{N}(u)}^{l}$ and its past hidden state $h_{u}^{l-1}$, as in equation \eqref{item-update}. While the representation of the rated items $h_{\mathcal{N}(u)}^{l}$ is updated aggregating the information from the linking edges $h_{uw}^{l}$ and the item nodes $h_{w}^{l-1}$, as in equation \eqref{N(U)-update}. Then, the item node $w$ is updated respectively as shown in the following equations: 
    \begin{align}
		h_{uw}^{l}&= \sigma(W_{E}^{l} \cdot \text{AGG}_{E}(h_{uw}^{l-1},h_{u}^{l-1},h_{w}^{l-1}))   \label{edge-update},\\
		  h_{\mathcal{N}(u)}^{l}&=\sigma(W_{\mathcal{N}(U)}^{l} \cdot \text{AGG}_{U}(\text{AGG}_{UW}(h_{w}^{l-1},h_{uw}^{l})))\label{N(U)-update},\qquad &\forall w \in \{\mathcal{N}(u)\},\\
		h_{u}^{l}&=\text{concat} (W_{U}^{l} \cdot h_{u}^{l-1},h_{\mathcal{N}(u)}^{l} ) \label{item-update},\\
		 h_{N(w)}^{l}&=\sigma(W_{\mathcal{N}(W)}^{l} \cdot \text{AGG}_{W}(\text{AGG}_{WU}(h_{u}^{l-1},h_{uw}^{l}))),\qquad &\forall u \in \{\mathcal{N}(w)\},\\
		h_{w}^{l}&=\text{concat} (W_{W}^{l-1} \cdot h_{w}^{l},h_{N(w)}^{l} ), \label{w-update}
\end{align}
    where $e_{uw}$ represents the edge representation for edge linking node $u$ and $w$, $h_{uw}^{l},h_{u}^{l}, h_{w}^{l}, h_{\mathcal{N}(u)}^{l}, h_{\mathcal{N}(w)}^{l}$ are hidden states at $l^{th}$ layer, $W_{E}^{l}, W_{U}^{l},W_{\mathcal{N}(U)}^{l},  W_{N(W)}^{l}, W_{W}^{l}$ are trainable weight matrices at $l^{th}$ layer, $\text{AGG}_{E}(\cdot)$, $\text{AGG}_{U}(\cdot), \text{AGG}_{UW}(\cdot)$, $\text{AGG}_{W}(\cdot), \text{AGG}_{WU}(\cdot)$ are user-chosen aggregation functions, $\sigma(\cdot)$ is the activation function, and $\text{concat}$ denotes the vector concatenation.

    There exist many modeling methodologies for bipartite graphs that could fit into the above framework. For instance, \citet{Zhang_Fraud} propose a similar model structure as the framework and apply the attention mechanism as the aggregation function, since different items may have different impacts when learning uses' representations. Instead of using all neighbors, \citet{Li_spam} utilize a sampling technique when aggregating neighbors' information in each iteration. There are also literature using the above framework as a building block and combining clustering methodology to learn a hierarchical representation of the graph, since hierarchical representation with various GNN models could achieve satisfactory performance. For instance,  \citet{Li_hierarchical} utilize the node embedding generated from the framework to cluster users into different communities and make a recommendation based on both community information and user information. Specifically, the user information is decomposed into two orthogonal spaces representing community-level information and individualized user preferences. \citet{Li_bipartite} treat the framework as a GNN module and stack it in a hierarchical fashion. With the embedding generated from the framework, clustering algorithms are performed to generate a coarsened graph which is used as an input for the next GNN layer.

\subsection{Multi-relation graph}
Instead of a simple homogeneous graph where all nodes and edges have the same type, in real life, there may exist multiple relations between nodes. For example, in a malicious account detection task, we could construct an Amazon review network. Users are the nodes in the graph and there are three relations between users: reviewing the same product, having the same star rating, and having similar texts. The multi-relation graph is denoted as $\mathcal{G}$ : $\mathcal{G}=(\mathcal{V},\mathcal{E}_{1:R})$, where $\mathcal{V}$ is the set of nodes, $\mathcal{E}_{1:R}$ is the set of edges, $R$ is the number of node types, $e_{i,j}^r\in \mathcal{E}_{r}$ is an edge between node $i,j$ with a relation $r\in \{1,...,R\}$.

A frequently used approach is to transform the heterogeneous graph into multiple homogeneous graphs by extracting $R$ subgraphs \{$\mathcal{G}^r=(\mathcal{V},\mathcal{E}_r), r=1,...,R$\} from it. Each subgraph $\mathcal{G}^r $only preserves one type of edge and thus is homogeneous. This line of work could be unified in a two-step framework. The first step implements the sub-graph aggregation to aggregate neighbor information in each sub-graph as shown in equation \eqref{within-relation}. The second step is to conduct the inter-relation aggregation to aggregate relation-specific embeddings as equation \eqref{inter-relation},
\begin{align}
		h^{l}_{i,r}&=f( \text{AGG}_{r}\{h^{l-1}_{j,r}\}) \label{within-relation}, \qquad &\forall j \text{ s.t.} (i,j)\in\mathcal{E}_r,\\
		h^{l}_{i}&=g( \text{AGG}\{h^{l}_{i,(1:R)},h^{l-1}_{i}\}) \label{inter-relation},
\end{align}
	where $h^{l}_{i,r}$ is the subgraph-specific embedding of node $i$ in subgraph $r$ in $l^{th}$ layer, $h^{l}_{i}$ is the general embedding of node $i$ in $l^{th}$ layer, $\text{AGG}_r(\cdot)$ is the aggregation function in subgraph $r$, $\text{AGG}(\cdot)$ is the inter-relation aggregation function, and $f(\cdot),g(\cdot)$ are user-defined functions.

There are multiple methods that could be categorized into the above two-step framework. For instance, in a fraud classification task, \citet{Liu_18} observe that fraudsters tend to congregate in topology and thus use weighted sum for within-relation aggregation to capture this congregation pattern. They then apply an attention mechanism for inter-relation aggregation to learn the significance for each sub-graph as follows:
\begin{align*}
	h^{l}_{i,r}&=\sigma( \text{Weighted mean}\{h^{l-1}_{j,r}\}),\qquad &\forall j \text{ s.t.} (i,j)\in\mathcal{E}_r,\\
	h^{l}_{i}&=\sigma(X_i W+\text{Attention}\{h^{l}_{i,(1:R)}\}),
\end{align*}
where $X_i$ is the feature vector for node $i$, $W$ is a trainable matrix, $\sigma(\cdot)$is the activation function, and $\text{Attention}$ denotes the attention aggregator.

To incorporate neighbors' information, \citet{Dou} use the mean aggregator for within-relation aggregation. To reduce the computational cost and keep the relational importance information, they apply a pre-calculated parameter $p_r^l$ as the weight in the intra-relation aggregation step. They also test several aggregating functions when aggregating relation-specific embeddings with the following structure:
\begin{align*}
	h^{l}_{i,r}&=\sigma( \text{Mean}\{h^{l-1}_{j,r}\}), \qquad &\forall j \text{ s.t.} (i,j)\in\mathcal{E}_r,\\
	h^{l}_{i}&=\sigma(h^{l-1}_{i}+\text{AGG}\{ h^{l}_{i,r}\cdot p_r^l\}),  \qquad &\forall r\in (1,...,R),
\end{align*}
where $p_r^l $ is a pre-trained weight.

 Since different relations provide various facets of user characteristics, relationship-specific embedding may have different statistical properties, which may cause trouble when aggregating them in a lower-level space. To deal with that, \citet{Wang} project the relation-specific node embedding to higher spaces using multi-layer perception (MLP) and concatenate them with relation-level attention:
\begin{align*}
	h_{i,r}^l&= \text{MLP}\{h_{i,r}^{l-1}\},\text{where }h_{i,r}^1=\text{Attention}\{x_{j,r}: \forall j \text{ s.t.} (i,j)\in\mathcal{E}_r\}, \\
	h^{l}_{i}&=\text{Concatenation with attention}\{h^{l}_{i,(1:R)}\},
\end{align*}	
where $\text{Attention}$ denotes the attention aggregator.
\subsection{Dynamic graph}
The previously mentioned neural network models generally focus on a static graph. However, in real-life settings, a graph may be dynamically evolving since relations may be updated with time. For example, in order to predict the loan default risk, a guarantee network needs to be updated, adding  newly-constructed guarantee relationships and removing companies that have fully paid the loan. With the rapid development of graph neural network methodologies on static graphs, there emerges a trend to extend GNN models to a dynamic setting. Consider a sequence of $T$ graphs $\mathcal{G}^{seq}=\{\mathcal{G}_1,...,\mathcal{G}_T\}$, where $\mathcal{G}_i=(\mathcal{V}_i,\mathcal{E}_i)$ represent the graph at $i^{th}$ time point. The feature matrices are represented as $X=\{\textbf{X}_1,...,\textbf{X}_T\}$ and the adjacency matrices are $A=\{\textbf{A}_1,...,\textbf{A}_T\}$. 

To capture the sequential pattern in the dynamic graph, a common approach is to train a GNN to generate the node embedding at each time stamp and then utilize a recurrent neural network to aggregate the information. For example, \citet{Cheng_IJCAI} obtain the node embeddings at each time step by training a GCN with multi-head attention. Then, they utilize the GRU to capture the sequential pattern with a temporal attention layer to capture the temporal variation over timestamps. Similarly, \citet{Yang} first aggregate node and edge information in each snapshot and then employ a LSTM operator to capture the temporal variations in the node enbeddings.

In the aforementioned methods, a graph neural network is learned for feature aggregation and an RNN model is trained to capture the sequential pattern of the node embeddings. However, in reality, a node may appear and disappear, which may worsen the performance of the RNN model when updating the node representation. In a guarantee network, for example, a company that has borrowed loans could disappear from the graph after it pays all the debts and could appear again backing up other companies' loans. To overcome the limitation, \citet{EvolveGCN} proposed EvolveGCN which utilizes a recurrent neural network to evolve the GCN parameters instead of updating the node embeddings. For each time point $t$, a GCN model is constructed as follows to fit the graph $\mathcal{G}_t$:
\begin{align*}
H^{l+1}_t=\sigma(\tilde{D}_t^{-\frac{1}{2}}\tilde{A_t}\tilde{D}_t^{-\frac{1}{2}}H^{l}_t W^{l}_t),
\end{align*}
where $\tilde{A}_t$ is the adjacancy matrix with added self-connections, $\tilde{D}_t=\text{diag}(\sum_j \tilde{A}_{ij})$ is the degree matrix, $W^{l}_t$ is trainable weight matrix of $l^{th}$ layer, $H^{l}_t$ is the matrix of activation in the $l^{th}$ layer, and $\sigma(\cdot)$ is the activation function.

To update the weight matrix $W^{l}_t$, \citet{EvolveGCN} propose two methods. The first method considers $W^{l}_t$ as a hidden state of the dynamics and update it using a GRU model, as shown in equation \eqref{evolve1}. The second method treats $W^{l}_t$ as an output state which is updated using a LSTM method, as shown in equation \eqref{evolve2}. The structure for both methods is as follows:
\begin{align}
W^{l}_t&=\text{GRU}(H_t^l,W^{l}_{t-1}) \label{evolve1},\\
W^{l}_t&=\text{LSTM}(W^{l}_{t-1}).
\label{evolve2}
\end{align}
Compared to the second method, the first method incorporates the updated node embedding in the recurrent neural network and it may lead to better performance when node features are informative.

\section{Application}
\label{session5}
{\color{black} In this section, we have detailed some financial applications that the GNN methods have been commonly applied on. We have also summarized features, graphs, methods, evaluation metrics, and baselines used in each financial application in the supplementary materials.}

\subsection{Stock movement prediction}
Though there are still debates on whether stocks are predictable, stock prediction receives great attention and there are rich literature on predicting stock movements using machine learning methods. However, the task of stock prediction is challenging due to the volatile and non-linear nature of the stock market. Traditionally, there are two major approaches to handle the task: technical analysis and fundamental analysis \citep{Sawhney}. Technical analysis utilizes numerical features such as closing prices and trading volumes, while the fundamental analysis approach includes non-numerical information, such as news and earning calls. The limitation of these non-graph approaches is that they often have a hidden assumption that the stocks are independent. To take the dependence into account, there is an increasing trend to represent the stock relations in a graph where each stock is represented as a node and an edge would exist if there are relations between two stocks. Predicting multiple stocks' movements could then be formed as a node classification task and GNN models could be utilized to make the prediction.

  In this section, we summarize the literature on stock prediction applying GNN methods, where table \ref{stock} presents the key features and graphs used in this application. There also exist challenges to apply the GNN methods in the stock prediction task. Unlike other fields where the benchmark graphs are available, to the best of our knowledge, there is no off-the-shelf graph representing inter-stock relations. With abundant relations existing in the financial system, it becomes challenging to obtain and select the relation for graph construction. Moreover, owing to the volatility of the stock market, how to model the sequential features and capture the temporal patterns are also critical. Also, the financial industry has rich data sources including financial statements, news and pricing information, which impose difficulty on modeling the data.

There are multiple ways to construct the stock relational graph. For instance, believing that correlation on historical prices reflects the inter-stock relation, \cite{Li} construct the graph using the correlation matrix of historic data to predict the movement of Tokyo stock price index. On the other hand, \citet{Matsunaga} borrow information from knowledge bases and construct supplier, customer, partner, and shareholder relational graphs. With multiple ways of graph construction, there doesn't exist a "best" graph due to the lack of graph evaluation methods. Future work could be done to design a graph evaluation method to help researchers better construct a relational graph.

To effectively process the sequential data and incorporate related corporations' information \citet{Chen} propose a joint model using LSTM and GCN to predict the stock movement. However, \citet{Chen}'s approach assumes that the relations between stocks are static, which may not reflect the reality. Instead, \citet{Feng} propose a temporal graph convolution layer to capture the stock relations in a time-sensitive manner, so that the strength of relation could be evolving over time. The relations are then updated based on historical pricing sequences and the proposed method obtained better performance compared to GCN. Believing that stock description documents also contain information reflecting the changes in companies' effect, \citet{Ying} capture the temporal relation by both sequential features and stock document attributes with a time-aware relational attention network.

The aforementioned methods focus on capturing the temporal dependencies, while \citet{Sawhney} focus on fusing data from different sources. \citet{Sawhney}  propose a multipronged attention network to jointly learn from historical price, social media, and inter stock relations. Encoded pricing and textual information are used as node feature inputs to GAT, where the graph information comes from the Wiki company-based relations. The attention mechanisms is applied to allocate different weights on various data sources and latent correlations may learned via the attention layers.

\subsection{Loan default risk prediction}
For commercial banks and financial regularity institutions, monitoring and assessing the default risk is at the heart of risk controlling process. As one of the credit risks, default risk is the probability that the borrower fails to pay the interest and principal on time. With a binary outcome, loan default prediction could be seen as a classification problem and is commonly addressed utilizing user-related features with classifiers including neural network \citep{nn} and gradient boosted trees \citep{gbm}. Since the probability that a borrower defaults may be influenced by other related individuals, there is plenty of literature forming a graph to reflect the interactions between borrowers. With the rapid growth of GNN methods, GNN methods are widely applied on the graph structure for loan default predicting problems. There are currently three lines of work focusing on various types of loans: guarantee loans, e-commerce loans, and other loans.

The guarantee loan allows small entrepreneurs to back each other in order to increase their credibility. It has a debt obligation contract that specifies that if one corporation fails to pay the debt, its guarantor needs to pay for it. A guarantee network naturally arises where each node is a company and directed edges represent the guarantee relationship. To learn a better representation of the network, \citet{Cheng_network_loan} utilize the graph attention layer and design a objective function, so that vertices with similar structures will be closer in the learned feature space. Since the guarantee relations changes with time, \citet{Cheng_IJCAI} forms a dynamic guarantee network to represent the dynamics. A recurrent graph neural network layer is developed to learn the temporal pattern and attentional weights are learned for each time point via an attention architecture. 

Unlike guarantee loans that the loan information could be  naturally represented in a directed graph, other loan types may not have a clear graph structure and researchers need to construct the graph based on interactive information. For instance, \citet{Xu} construct a user relation graph where users are connected by various relationships, such as social connections, transactions, and device usage. However, the interactive graph may also contain noisy data, which may be irrelevant. Since the massive interactive information may be noisy and the impacting supply-chain information is deficient, \citet{Yang} extract supply-chain relations while predicting loan defaults. Forming the interaction data as a graph, \citet{Yang} formulate the supply chain mining task as a link prediction task and thus construct a supply chain network, which is then used to predict the default probability with GNN methodology.

With the rise of e-commerce, e-commerce consumer lending service is gaining popularity to enhance consumers' purchasing power. Able to obtain information from multiple facets, the e-commerce platform could have multi-view data and multi-relation networks, which may require sophisticated modeling methodology. For instance, in order to predict the default probability for each consumer with multi-view data, \cite{Liang_WSDM} utilize a hierarchical attention mechanism to encode the features on each view. Exploring multiplex relations, \citet{Hu} propose an attributed multiplex graph-based model with relation-specific layer and attention mechanism to jointly model multiple relations. To simultaneously model the labeled and unlabeled data, \cite{Wang} proposed a semi-supervised graph neural network approach and obtain interpretable results.

\subsection{Recommender system of e-commerce}
With the rapid growth of e-commerce, customers gradually get used to shopping online and are exposed to a numerous range of products. To alleviate the burden of users choosing the appropriate item, the recommender system is developed to suggest products to users based on predicted item ratings. Presenting the user and item information in a graph, GNN methodologies are widely used in recommender system related tasks including click rate prediction and fake review detection. 

To accurately predict users' preferences and recommend the appropriate item, it is vital to exploit information for users, items, and their interactions. A widely-used representation of that information is a bipartite graph, where nodes are of two types, user and item, and edges represent that there are relationships between user and item nodes. Noticing that the community the user belongs to may affect the shopping decision, for example, a user belongs to the traveler group may purchase travel-related items, \cite{Li_hierarchical} combine the bipartite graph modeling algorithm with clustering techniques to reflect the community impact and the individual preference. However, \cite{Li_hierarchical}'s approach only considers the hierarchy in the user side, while items may also have hierarchical information. To fill the gap, \cite{Li_bipartite} stack several GNN modules hierarchically to capture the hierarchical structure in both user and item perspectives. Embeddings learned from the GNN layer are clustered and used as an input for the next GNN layer, which could preserve the high-order hierarchical connections. 

The recommender system is mainly based on the past history of the user, including its rating and reviews on the item. However, fake ratings and feedback may be posted by the fraudsters to seek financial benefits. To detect fraudulent reviews on the e-commerce platform, \citet{Kudo} construct a directed and signed comment graph with a signed graph convolutional network approach. Compared to \citet{Kudo}'s approach which only considers the comment graph, \citet{Li_spam} integrate both the bipartite user-item graph and a comment graph to capture the local and global context of the comments. Noticing that camouflage behaviors of the fraudsters may deteriorate the performance of fraud detection mechanism and has seldom been considered by prior works, \citet{Dou} propose a model against both feature and relation camouflage. For each node, only informative neighbors are selected for the next aggregation step, utilizing a similarity measure and a reinforcement learning mechanism. While the above literature focus on the fraud review detection side, there are also works that accomplish both fraud review detection and item recommendation tasks. For example, \citet{Zhang_Fraud} propose a GCN-based framework that performs both item recommendation and fraud detection in an end-to-end manner, while each of the tasks is beneficial for the other one.

\subsection{Fraud detection}
\label{fraud}
Including payment fraud, identity theft, financial scam and insurance fraud, financial fraud has a variety of types and has an increasing trend \citep{Kurshan}. Observing that fraudsters tend to have abnormal connectivity with other users, there is a trend to present users' relations in a graph and thus, the fraud detection task could be formulated as a node classification task.

Aiming to detect the malicious accounts, who may attack the online services to seek excessive profits, \citet{Liu_18} find out that fraudsters have two patterns: device aggregation and activity aggregation. Due to economic constraints, attackers tend to use limited number of devices and perform activities in a limited time, which may be reflected in the local graph structure. With this observation, \citet{Liu_18} propose a variant of GCN and use sum operators to capture the aggregation pattern. While the malicious accounts may be aggregated together, \citet{Liu_19} argue that the normal accounts could also be connected with the malicious account and may be mislabeled as malicious, which adds noisy signals to the graph. Taking into account that the graph could be noisy and nodes may have different impact, \citet{Liu_19} propose an adaptive path layer to adaptively select the important neighbor nodes that contribute most to the target node. Applying the GNN methodologies proposed by \citet{Liu_18}, \citet{Liang} stack multiple adaptive path layers to aggregate neighbors' features and have great performance in a insurance fraud detection task.

The above literature focus on a bipartite graph, where nodes are either account or devices, and there are also literature utilizing other types of graphs. For instance, \citet{Jiang_anomaly} construct a homogeneous user network connecting user nodes based on their similarity of behaviors and apply the GCN model. \citet{Rao} focus on a dynamic graph of registration records and implement GCN layers on structural and temporal subgraphs. 

Besides the literature developing GCN-based methodologies, there are also fraud-detection related literature applying GNN models with other structures. Since GCN may suffer from over-smoothing problem and have shallow model structure, \citet{Lv} propose to replace the graph convolutional matrix with auto-encoder to increase the depth of the neural network. \citet{Zhao_loss} argues that GNN models with random-walk based losses have poor performance in anomaly detection task, since nodes of the same label may not be closer. They then propose a loss function which leads to better performance and has bounded prediction error.

\subsection{Event prediction}
Financial events, including revenue growth, acquisition and bankruptcy, could provide valuable information on market trends and could be used to predict future stock movement. Therefore, it draws great attention on how to predict next financial event based on past events and currently GGNN model is often used to accomplish the task. For example, given a sequence of financial events of Chinese listed companies, \citet{Yang_event} aim to predict the next event type and construct an event graph where each node is a financial event and edges are weighted using the frequency of the event pairs. \citet{Harl} transform a binary classification problem into an event prediction task by dividing the loan application process into several events and predicting whether the next event would be accepting or rejecting the application. Utilizing the GGNN model, they obtain the predictions with high accuracy.

\section{Challenges}
\label{session6}
\subsection{Graph evaluation methods}
To justify the inclusion of a graph, a commonly-used evaluation method is to compare the outcome for a graph-based machine learning method with a graph-free machine learning method \citep{Feng,Li}. However, there is little discussion on comparing different graphs' effects and quality, while existing literature discussing graph comparisons are often inadequate. For example, \citet{Liang} visualize the structural patterns in different graphs, concluding that the device sharing graph is more appropriate based on the observed patterns. Without presenting the evaluation metric for each graph, the graph comparison based on the visualization may not be adequate. The problem may be more severe in similarity-based graph construction since a threshold needs to be set to determine whether an edge exists. Different threshold values may lead to completely different graphs and thus affect the model performance. Thus, justification on threshold setup during graph construction is of great importance. Some efforts have been made to tackle this problem, such as utilizing a reinforcement learning approach to automatically select the optimal threshold  \citep{Dou}. More attention may need to be drawn to develop a framework to assess the graph quality systematically.

\subsection{Explainability}
Combining both graph structural information and feature information, GNN models are often complicated and it is challenging to make an interpretation. Recently, there are some literature focusing on the explainability of GNN models. GNNExplainer \citep{GNNExplainer}, for example, is proposed to provide an interpretable explanation on trained GNN models such as GCN and GAT. Model explainability in financial tasks is of great importance, since understanding the model could benefit decision-making and reduce economic losses. However, there is little literature studying the explainability of GNN models in a financial application, which often accompanies by heterogeneous and dynamic graphs. Current literature focuses on relatively simpler graphs. For example, \citet{Li_explain} extend the GNNExplainer to a weighted directed graph and apply it on a Bitcoin transaction graph. \cite{Rao1} propose an explainable fraud prediction system that could operate on heterogeneous graphs consisting of different node and edge types. More work could be done on the explainability of GNN models with edge-attributed graphs and dynamic graphs, which are not yet considered.

\subsection{Task type}
Tasks of GNN models are commonly classified into three categories: node-level task, edge-level task, and graph-level task \citep{Wu}. However, the GNN methods applied on financial applications are mostly focused on the node-level task. For example, the stock movement prediction task is often formulated as a node classification task, where stocks are represented as nodes. There exist some literature focus on other types of task, for instance, \citet{Yang} aim to  mine the underlying supply chain network and formulate it as a link prediction task, but this line of work is rare. Since there are rich literature developing GNN methodologies on edge prediction or graph classification tasks, there are rich opportunities on applying recently-developed GNN methods on financial fields if financial tasks could be formulated into an edge or graph level task.

\subsection{Data availability}
To pursue reproducibility, it is a common practice to release both data sources and codes when publishing the paper. However, in the reviewed papers, only about 24\% of them release the code. Since the financial tasks are commonly based on real-world problems, data may have some restrictions due to the privacy obligation of the related corporations. Lack of open-source codes and datasets, it is hard to reproduce previous works and compare their methodologies in the latter literature. Thus, it is of great value to construct benchmark datasets, so that methods could be compared on the same data.

\subsection{Scalability}
In the real-world financial scenario, commercial data are often of large scales. For instance, \citet{AliGraph} utilize  data from a popular e-commerce platform and it contains about 483 million nodes with 231 million edges. How to improve the scalability of GNNs is vital but challenging. Computing the Laplacian matrix becomes hard with millions of nodes and for a graph of irregular Euclidean space, optimizing the algorithm is also difficult. Sampling techniques may partially solve the problem with the cost of losing structural information. Thus, how to maintain the graph structure and improve the efficiency of GNN algorithms are worth further exploration. 

\newpage
\section*{Supplementary material}
In the supplementary materials, we present materials that are not covered in the main text. The supplementary materials contain the summary table for each financial application, figures categorizing major GNN methodologies for each graph type and acronyms used in the text.

\subsection*{Summary tables for financial applications}

\begin{longtable}{p{.1\textwidth}p{.25\textwidth}p{.15\textwidth}p{.1\textwidth}p{.1\textwidth}p{.15\textwidth}}
		\toprule
		Reference & Feature \footnote{{\color{black}The red color format represents the acronym format and the full form could be found in the supplementary material section by clicking the acronyms.}} & Graph & Method &Evaluation metric& Baseline method\\
		\midrule
		\citet{Chen}  &7-day open, high, low, close prices and volume for \acrshort{CSI} listed companies& Corporation shareholder network & LSTM + GCN & Accuracy & \acrshort{LSTM}, \acrshort{GCN} \\
		\citet{Feng} & 5/10/20/30 days moving average of close price for \acrshort{SP500}/\acrshort{NYSE} listed companies &Wiki company-based relation network \& sector-industry  stock relation network& \acrshort{RSR} & \acrshort{MSE}, \acrshort{IRR}, \acrshort{MRR} & \acrshort{SFM}, \acrshort{LSTM}, \acrshort{GBR}, \acrshort{GCN}\\
		\citet{Matsunaga} & 5/10/20/30 days moving average of close prices for \acrshort{Nikkei 225} listed companies &Supplier, customer, partner and shareholder  relation graph & \acrshort{RSR} &Return ratio, Sharpe ratio & Model with subsets of relations\\
		\citet{Ying} & 5/10/20/30 days moving average of close prices and stock description documents for \acrshort{SP500} or \acrshort{NYSE} listed companies &Wiki company-based relation network & \acrshort{TRAN} & \acrshort{MSE}, \acrshort{MRR}, \acrshort{IRR} & \acrshort{RSR}, \acrshort{GCN}, \acrshort{LSTM}\\
		\citet{Li}&News of \acrshort{TPX} 500/100 listed companies &Stock correlation graph & \acrshort{LSTM-RGCN} &Accuracy & Naïve Bayes, \acrshort{LR}, \acrshort{RF}, \acrshort{HAN}, transformer, \acrshort{S-LSTM} \\
		\citet{Sawhney}&Price and social media information for companies listed in the \acrshort{SP500} index  or \acrshort{NYSE} or \acrshort{NASDAQ} markets&Wiki company-based relation network & \acrshort{MAN-SF} & Accuracy, F1, \acrshort{MCC} & \acrshort{ARIMA}, \acrshort{RF}, \acrshort{TSLDA}, \acrshort{HAN}, StockNet, LSTM+GCN\\
		\citet{Sawhney_audio}&Text and audio features of earning calls for companies in the \acrshort{SP500} index & Stock earning call graph & \acrshort{VolTAGE} &\acrshort{MSE}, R-squared& \acrshort{LSTM}, \acrshort{HAN}, \acrshort{MDRM}, \acrshort{HTML}\\
		\citet{Liou}&News and attributes for stock tags&News co-occurance graph &\acrshort{HAN1} &Accuracy, \acrshort{MCC} & \acrshort{RF}\\
		\bottomrule
	\caption{Summary of stock movement prediction literature}
	\label{stock}
\end{longtable}
\begin{longtable}{p{.1\textwidth}p{.225\textwidth}p{.15\textwidth}p{.1\textwidth}p{.1\textwidth}p{.175\textwidth}}
		\toprule
		Reference & Feature & Graph & Method &Evaluation metric& Baseline method\\
		\midrule
		\citet{Wang}  & / & Multiple user networks & \acrshort{SemiGNN} & \acrshort{AUC}, \acrshort{KS}& \acrshort{Xgboost}, \acrshort{LINE}, \acrshort{GCN}, \acrshort{GAT}\\
		\citet{Cheng_network_loan}  &Active loan behavior, historical behavior, and user profile & Guarantee network & \acrshort{HGAR} & \acrshort{AUC}, \acrshort{precision@k} & \acrshort{GF}, \acrshort{DW}, node2vec, \acrshort{AANE}, \acrshort{SNE}, \acrshort{GAT}\\
		\citet{Cheng_KDD}& Customer profile, loan information, guarantee profile, and loan contract &Guarantee network & \acrshort{TRACER} & F1, \acrshort{precision@k} & \acrshort{LR}, \acrshort{GBDT}, \acrshort{DNN}\\
		\citet{Cheng_IJCAI} &Loan behavior and company profile &Temporal guarantee network & \acrshort{DGANN} & \acrshort{AUC} & \acrshort{GF}, \acrshort{GCN}, node2vec, \acrshort{GAT}, \acrshort{SEAL}, \acrshort{RNN}, \acrshort{GRNN}\\
		\citet{Yang} &Credit-related features, spatial features, and temporal features &Temporal smallbuiness en-trepreneur network & \acrshort{ST-GNN} &\acrshort{AUC}, \acrshort{KS} & \acrshort{GBDT}, \acrshort{GAT}, \acrshort{STAR}\\
		\citet{Hu_Loan} &User credit exposure features &Alipay user and applet graph & \acrshort{AMG-DP} &\acrshort{AUC}, \acrshort{KS} & \acrshort{MLP}, \acrshort{Xgboost}, node2vec, \acrshort{GraphSAGE}, \acrshort{GAT}, \acrshort{HAN}, \acrshort{SemiGNN} \\
		\citet{Liang_WSDM} & Device information and trading features&User relationship graph & \acrshort{MvMoE} &\acrshort{AUC}& \acrshort{GBDT} \\
		\citet{Xu} & User profile and transaction summary &Auto loan network & \acrshort{GRC} & Precision, recall, F1 & \acrshort{SVM}, \acrshort{MLP}, \acrshort{GCN} \\
	    \citet{Lee} &Loan information, credit history, and soft information  & Borrower’s relation network & / & Accuracy, precision, recall, F1, \acrshort{AUC} & \acrshort{SVM}, \acrshort{RF}, \acrshort{Xgboost}, \acrshort{MLP}, \acrshort{GCN}\\
		\bottomrule
	\caption{Summary of loan default risk prediction literature}
	\label{loan}
	\end{longtable}

\begin{longtable}{p{.1\textwidth}p{.225\textwidth}p{.175\textwidth}p{.1\textwidth}p{.1\textwidth}p{.15\textwidth}}
		\toprule
		Reference & Feature & Graph & Method &Evaluation metric& Baseline method\\
		\midrule
		\citet{Li_spam}  & Features for item, user and comment & Xianyu comment graph & \acrshort{GAS} &  \acrshort{AUC}, F1, recall& \acrshort{GBDT}\\
		\citet{Zhang_Fraud}  &Behavioral features  & Yelp review network, Amazon review network & \acrshort{GraphRfi} & \acrshort{MAE}, \acrshort{RMSE}, precision, recall, F1 & \acrshort{RCF}, \acrshort{GCMC}, \acrshort{PMF}, \acrshort{ICF}, \acrshort{MF} \\
		\citet{Dou}  &Behavioral features  & Yelp review network, Amazon review network & \acrshort{CARE-GNN} &  \acrshort{AUC}, recall & \acrshort{GCN}, \acrshort{GAT}, \acrshort{RGCN}, \acrshort{GraphSAGE}, \acrshort{SemiGNN}\\
		\citet{Kudo}  &Behavioral features  & Amazon review network & \acrshort{GCNEXT} &  \acrshort{AUC} &  \acrshort{RGCN}, \acrshort{SIDE}, \acrshort{SGCN}\\
		\citet{Li_hierarchical}  & Purchasing power, number of transactions, item category, and shipping costs & E-commerce user-item network & \acrshort{Bi-HGNN} & Accuracy, \acrshort{AUC}, F1 & \acrshort{GraphSAGE}, \acrshort{Diffpool}\\
		\citet{Li_bipartite}  & Click and transaction logs  & Taobao user-item network & \acrshort{HiGNN} &\acrshort{AUC} & \acrshort{DIN}, \acrshort{GE} \\
		\bottomrule
		\caption{Summary of literature on recommendation system of e-commerce}
	\end{longtable}
\begin{longtable}{p{.1\textwidth}p{.2\textwidth}p{.175\textwidth}p{.1\textwidth}p{.1\textwidth}p{.175\textwidth}}
		\toprule
		Reference & Feature & Graph & Method &Evaluation metric& Baseline method\\
		\midrule
		\citet{Liang}  & Insurance claim history, shipping history and shopping history & Device sharing graph & / & F1, \acrshort{DE} & \acrshort{GBDT}, node embeddings\\
		\citet{Lv} & Purchase history of 1000 commodity categories & JD Finance anti-fraud graph &\acrshort{AutoGCN}  & \acrshort{AUC} &  \acrshort{GCN}, \acrshort{GAT}, GCN with attention\\
		\citet{Rao1} & Individual risk features & Transaction records graph & xFraud   & \acrshort{AUC} & \acrshort{LR}, \acrshort{DNN}, \acrshort{GAT}, \acrshort{GCN}, \acrshort{HGT} \\
		\citet{Zhu} & User features & Iqiyi  user network & \acrshort{HMGNN} &Precision, recall, F1, \acrshort{AUC} & \acrshort{LR}, \acrshort{Xgboost}, \acrshort{MLP}, \acrshort{GCN}, \acrshort{GAT}, \acrshort{ASGCN}, \acrshort{mGCN}\\
		\citet{Jiang_anomaly}&User behavioral features and content based features& Simulated user activity network &/ & Accuracy, precision, recall & \acrshort{RF}, \acrshort{SVM}, \acrshort{LR}, \acrshort{CNN}\\
		\citet{Liu_18}&User activities & Alipay one-month account-device network &\acrshort{GEM} & F1,  \acrshort{AUC}, precision-recall & Connected subgraph, \acrshort{GBDT}, \acrshort{GCN}\\
		\citet{Liu_19}&User activities & Alipay one-week account-device network &GeniePath &Accuracy & \acrshort{MLP}, node2vec, \acrshort{GCN}, \acrshort{GraphSAGE}, \acrshort{GAT}\\
	    \citet{Rao}&Registration profile and transaction features & Account-registration graph &\acrshort{DHGReg} & Precision & \acrshort{MLP}, \acrshort{GCN},  \acrshort{GAT}\\
	     \citet{Zhao_loss}&/ & Bitcoin-alpha graph &\acrshort{GAL} & Precision, recall, F1, \acrshort{AUC} & \acrshort{GCN}, \acrshort{GAT}, \acrshort{GraphSAGE}, \acrshort{DOMINANT}\\
		\bottomrule
		\caption{Summary of fraud detection prediction literature}
	\end{longtable}
\begin{longtable}{p{.125\textwidth}p{.1\textwidth}p{.2\textwidth}p{.1\textwidth}p{.175\textwidth}p{.15\textwidth}}
		\toprule
		Reference & Feature & Graph & Method &Evaluation metric& Baseline method\\
		\midrule
	    \citet{Harl} & Event features &Loan application event graph & \acrshort{GGNN} &Accuracy &/ \\
	    \citet{Yang_event} &Financial news   & Financial event graph & \acrshort{GGNN} & Accuracy, precision, recall, F1 & \acrshort{PMI}, \acrshort{DW}, \acrshort{LSTM}\\
		\bottomrule
	\caption{Summary of event prediction literature}
	\end{longtable}

\subsection*{Figures on GNN methods for each graph type}

\begin{figure}[H]
     \centering
     \begin{subfigure}[b]{\textwidth}
         \centering
         \includegraphics[width=\textwidth]{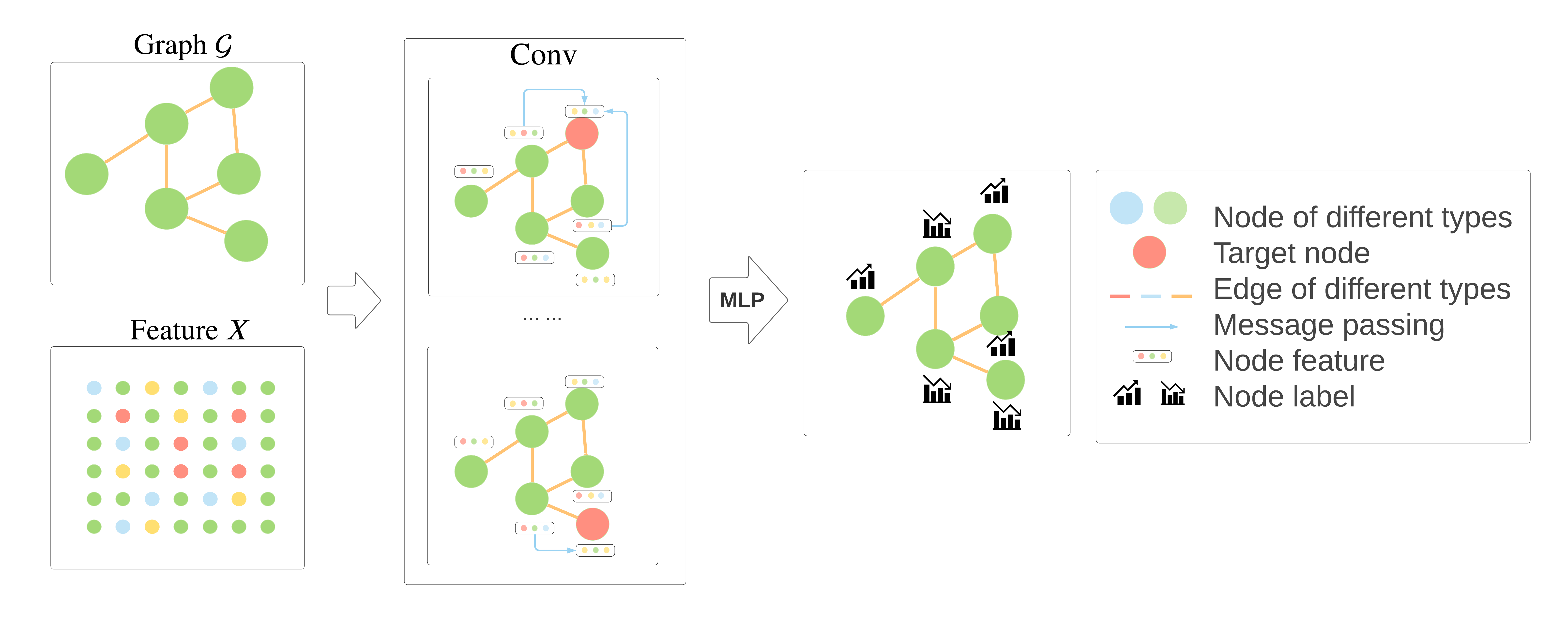}
         \caption{Graph neural network models for homogeneous graphs. With a graph adjacency matrix and a feature matrix as inputs, the graph convolutional process generates a representation for each target node by aggregating the information of its neighbors. A multi-layer perception layer is then applied to generate the predicted labels.}
     \end{subfigure}
\end{figure}
\begin{figure}[H]\ContinuedFloat
     \begin{subfigure}[b]{\textwidth}
         \centering
         \includegraphics[width=\textwidth]{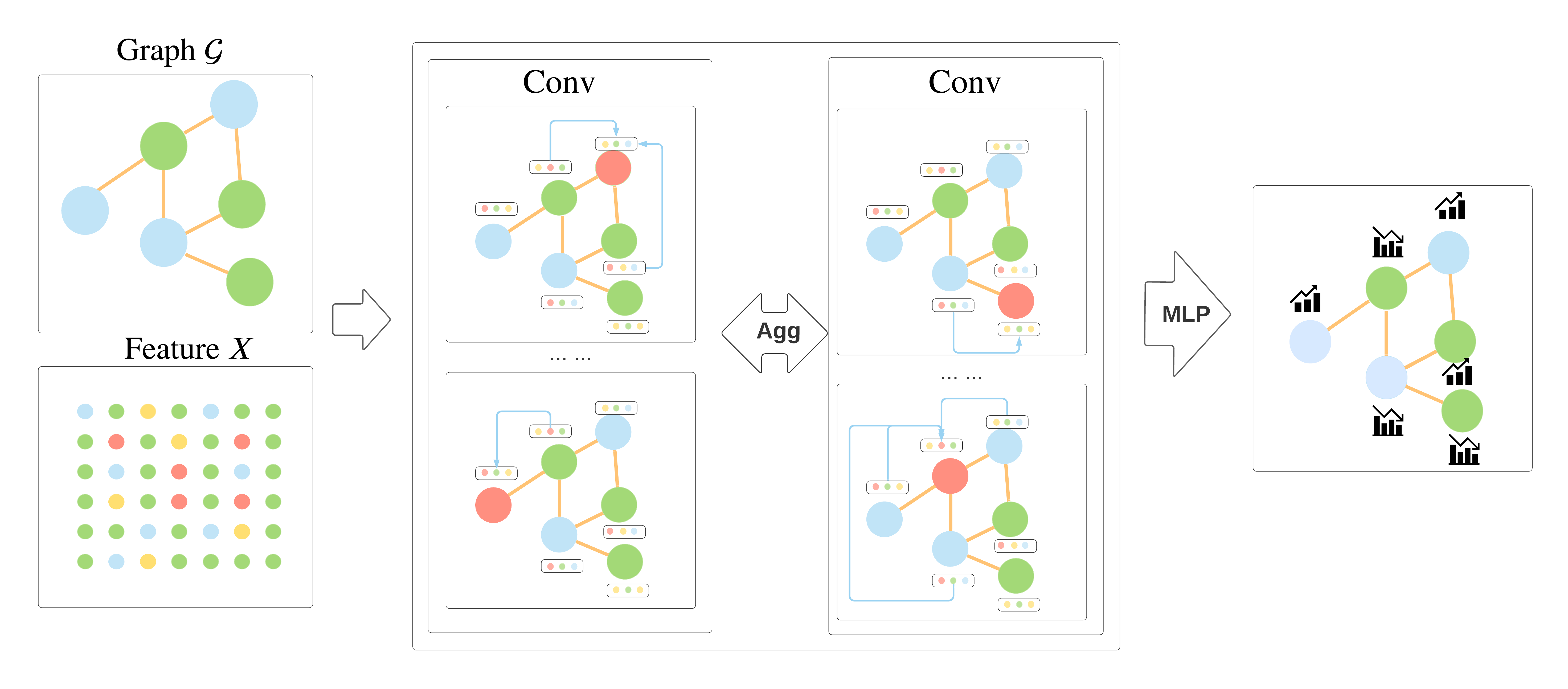}
         \caption{Graph neural network models for bipartite graphs. For nodes having the same type, the graph convolutional process updates their node representations using their own information and their neighbors' information. Aggregating functions are utilized to generate the node representations from the representations for both node types. A multi-layer perception layer is then applied to generate the predicted labels.}
     \end{subfigure}
\end{figure}
\begin{figure}[H]\ContinuedFloat
     \centering
    \begin{subfigure}[b]{\textwidth}
         \centering
         \includegraphics[width=\textwidth]{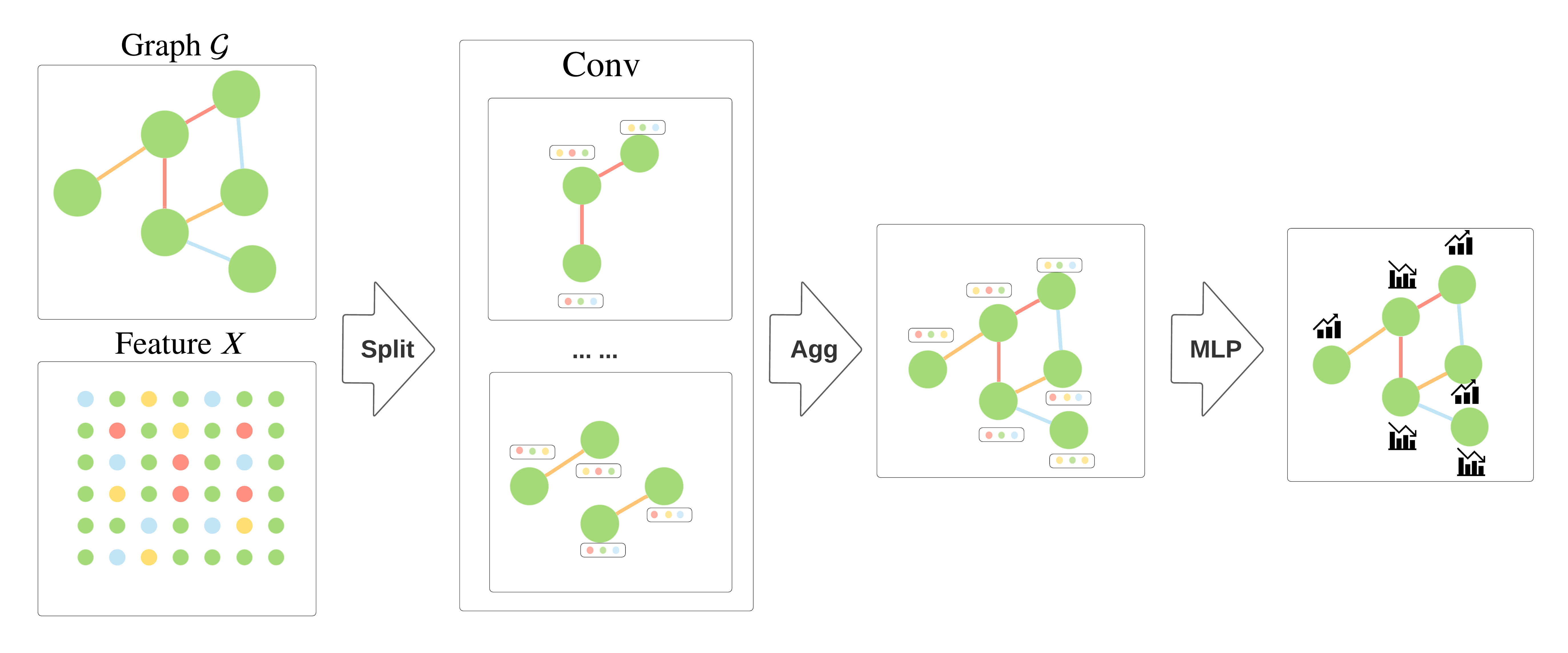}
         \caption{Graph neural network models for multi-relation graphs. A multi-relation graph is split into several sub-graphs in which edges are of the same type. Within each sub-graph, the graph convolutional process takes place to update the node representations. Then between-graph aggregation is utilized to obtain the final node representations. A multi-layer perception layer is then applied to generate the predicted labels. }
     \end{subfigure}
\end{figure}
\begin{figure}[H]\ContinuedFloat
     \begin{subfigure}[b]{\textwidth}
         \centering
         \includegraphics[width=\textwidth]{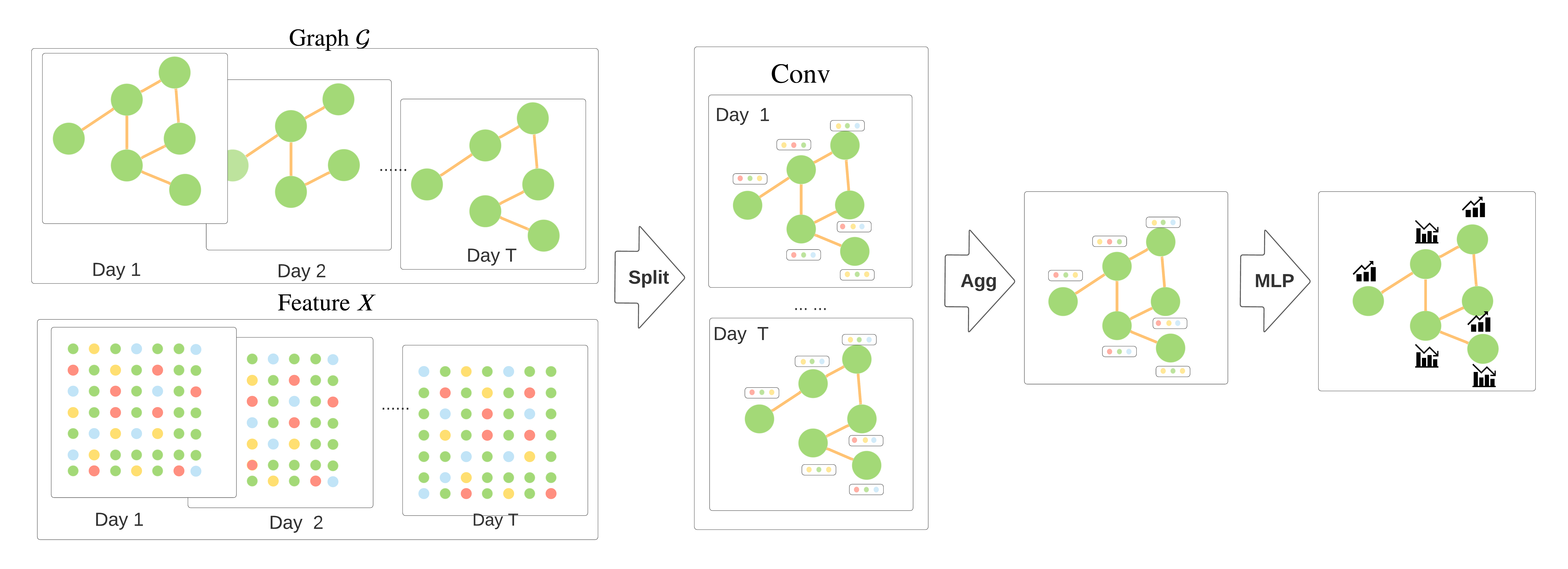}
         \caption{Graph neural network models for dynamic graphs.  A dynamic graph is often a sequence of graphs ordered by time. The graph convectional process generates node representations for graph at each time stamp. To get the most recent representation, node representations are aggregated using functions including the recurrent neural network. A multi-layer perception layer is then applied to generate the predicted labels.}
     \end{subfigure}
        \caption{Graph neural network models for different graph types. The term $\text{Conv}$ denotes graph convolution process. The term $\text{MLP}$ denotes the multi-layer perception. The term $\text{Agg}$ denotes the aggregation process. The term $\text{Split}$ denotes data splitting according to its characteristics.}
        \label{model}
\end{figure}
\printnoidxglossary[type=acronym,sort=letter]

\printnoidxglossary[sort=use]

\bibliographystyle{jds}
\bibliography{ref1}

\newpage

\end{document}